\documentclass{aa}  
\usepackage{multirow}
\usepackage{graphicx}

\usepackage{txfonts}

\usepackage[]{hyperref}
\hypersetup{unicode=true, colorlinks=true, linkcolor=[rgb]{0.0, 0.0, 1}, citecolor=[rgb]{0.0, 0.0, 1}, filecolor=[rgb]{0.0, 0.0, 1}, urlcolor=[rgb]{0.0, 0.0, 1}}
\usepackage{natbib}
\usepackage{scrhack} 
\usepackage{silence}
\begin{document}

 \title{Timing and spectral analysis of the 2025 outburst of 4U 1630$-$47 with \textit{NICER}}

       \author{Haifan Zhu
          \inst{1}\fnmsep\thanks{zhu@astro.rug.nl}
          \and
Mariano M\'endez\inst{1}\fnmsep\thanks{mariano@astro.rug.nl}
          \and
          Xiao Chen\inst{2}
          \and
          Wei Wang\inst{3}
          }

   \institute{Kapteyn Astronomical Institute, University of Groningen, P.O. BOX 800, 9700 AV Groningen, The Netherlands
    \and
    Department of Physics, College of Science, Southern University of Science and Technology, Shenzhen 518055, China 
    \and
             Department of Astronomy, School of Physics and Technology, Wuhan University, Wuhan 430072, China}

\date{Received XXX; accepted YYY}
 
\abstract
{We analyzed \textit{NICER} observations of the 2025 outburst of the black hole X-ray binary 4U~1630$-$47 to investigate the spectral--timing properties of its transient low-frequency quasi-periodic oscillations (QPOs) and millihertz-scale quasi-regular modulation (QRM). During the rising phase of the outburst, the QPO centroid frequency increased from $\sim 0.24$ Hz to $\sim 3.43$ Hz. Wavelet-based state separation shows that the with-QPO intervals are associated with a higher inner disk temperature and a lower \texttt{diskbb} normalization than the without-QPO intervals, while the photon index ($\Gamma$) shows weaker changes within the uncertainties. Near the outburst peak, the source displayed a weak QRM at $\sim 0.07$ Hz with a fractional rms amplitude of $\sim 4.7\%$, lower than that of the  heartbeat state observed in 2023. Phase-resolved Hilbert--Huang analysis shows that the inner disk temperature is positively correlated with the X-ray flux, the \texttt{diskbb} normalization is anticorrelated, and $\Gamma$ varies only weakly. Overall, the short-timescale spectral--timing variability is expressed most clearly through the disk-related parameters. The transient QPOs are therefore consistent with short-timescale disk-related variability during the rising phase, whereas the millihertz-scale QRM may represent a weaker heartbeat-like variability mode appearing near the outburst peak.}

   \maketitle

\section{Introduction}

Black hole X-ray binaries (BHXRBs) are systems in which a stellar-mass black hole accretes matter from a companion star. Most BHXRBs are transients, spending long intervals in quiescence and occasionally undergoing outbursts that last for weeks to months. During these episodes, the release of gravitational potential energy produces strong X-ray emission, making BHXRBs valuable laboratories for studying accretion under extreme conditions \citep{remillard2006x}.

The spectral and timing evolution of a typical outburst is often described in terms of a characteristic "q"-shaped track in the hardness-intensity diagram (HID), corresponding to transitions between the low-hard, hard-intermediate, soft-intermediate, and high-soft states \citep{belloni2005evolution}. Some systems, however, undergo so-called failed outbursts and remain in the hard state throughout the event \citep{capitanio2009failed,alabarta2021failed}. Quasi-periodic oscillations (QPOs) frequently appear during these states as narrow or moderately broad peaks in the power density spectrum (PDS). Since their discovery, low-frequency QPOs (LFQPOs) have been widely used as probes of the dynamics of the inner accretion flow and strong-gravity phenomena near the compact object \citep{motch1983simultaneous,van1989quasi,ingram2019review}.

Based on their spectral and timing properties, LFQPOs in black-hole binaries are commonly classified into type-A, type-B, and type-C categories. 
This classification is mainly based on the centroid frequency ($\nu$), quality factor $Q=\nu/\Delta\nu$, where $\Delta\nu$ is the full width at half maximum (FWHM), fractional rms amplitude, associated broadband-noise (BBN) component, and phase lag \citep{remillard2002characterizing,casella2005abc}. Current models are usually grouped according to whether they represent geometric scenarios, such as Lense--Thirring precession of the inner hot flow or jet-like structure \citep{schnittman2006precessing,ingram2009low,ingram2011physical,ma2021discovery}, or intrinsic scenarios, in which oscillations or instabilities develop within the accretion flow \citep{tagger1999accretion,chakrabarti1993smoothed,karpouzas2020comptonizing,bellavita2022vkompth,mastichiadis2022study,mastichiadis2026radiative}.

In addition to LFQPOs, some BHXRBs also show variability on very long timescales, typically $\sim 10$--$200$ s. In the light curve, this variability appears as quasi-regular flares or dips; in the PDS, this produces broad peaks at several to tens of millihertz. Such variability is often referred to as quasi-regular modulation (QRM) \citep{trudolyubov2001rxte,yang2022insight,zhao2023mhz}. A particularly striking example is the "heartbeat" state, in which the light curve displays electrocardiogram-like oscillations at millihertz frequencies \citep{neilsen2011physics}. This type of behavior has so far only been identified in a small number of systems, including GRS~1915 $+$105, IGR~J17091$-$3624, GRO~J1655$-$40, 4U~1630$-$47, and H1743$-$322 \citep{morgan1997rxte,remillard1999rxte,trudolyubov2001rxte,wang2024highly,altamirano2012low}. The best-known case is the $\rho$ class GRS~1915+105, whose quasi-periodic flares recur on timescales of $\sim 40$--$200$ s and are accompanied by strong spectral evolution within each cycle \citep{belloni2000model,neilsen2012radiation,weng2018statistical}. Phase-resolved studies have shown that the disk temperature increases with flux and the apparent \texttt{diskbb} normalization  decreases, a pattern commonly interpreted in terms of thermal-viscous or radiation-pressure-driven limit-cycle behavior in the inner disk \citep{lightman1974black,neilsen2012radiation}.

Most timing studies of BHXRBs are based on Fourier techniques \citep{miyamoto1991x,huang2018insight,zhu2023timing,yu2024timing}. In recent years, however, time-frequency methods have increasingly been used to study nonstationary variability, including wavelet analysis \citep{czerny2010model,ding2021qpos,chen2022wavelet,chen2022waveleta,jin2024wavelet} and the Hilbert--Huang transform (HHT; \citealt{yu2023hilbert,shui2024phase,shui2024recovery,shui2025phase}). These approaches are particularly useful when the oscillation is intermittent, evolves in frequency, or departs from strict periodicity. Wavelet analysis has been used, for example, to isolate QPO-active intervals and to compare the timing and spectral properties of different variability states in sources such as MAXI~J1535$-$571 and MAXI J1803$-$298 \citep{chen2022wavelet,chen2022waveleta,chen2024different,jin2024wavelet}. HHT, which is well suited to nonlinear and nonstationary signals \citep{Huang1998}, has been used to show that QPO peak broadening can be dominated by frequency modulation and to track phase-dependent spectral evolution in heartbeat-like states \citep{yu2023hilbert,shui2024recovery,shui2024phase}. Higher-order timing tools such as bicoherence have also provided useful information on nonlinear coupling between QPOs and BBN \citep{maccarone2002higher,maccarone2011coupling,arur2022using,hzhu2024bicoherence,zhu2024bicoherence}. 

4U~1630$-$47 is an active BHXRB in the Galaxy whose outbursts recur on timescales of roughly $\sim 600$--$700$ days \citep{jones1976uhuru,priedhorsky1986recurrent,parmar1995periodic,tanaka1996x}. The source lies toward the Galactic center and is heavily absorbed, with $N_{\rm H}\sim10^{23}\ \mathrm{cm^{-2}}$; strong optical extinction has so far prevented the identification of a secure optical counterpart and therefore any dynamical mass measurement \citep{gatuzz2019chandra,seifina2014black}. Spectral modeling nevertheless suggests a rapidly spinning black hole viewed at a relatively high inclination \citep{king2014disk,pahari2018astrosat,liu2022spins,kushwaha2023ixpe,seifina2014black,rawat2023detection}, 
and dust-scattering halo studies favor a distance of $\sim 11.5$~kpc \citep{kalemci2025dust}. 
The source is often observed in soft states and shows complex outflow behavior, including wind-related absorption features and jet activity reported in different epochs \citep{capitanio2015missing,pahari2018astrosat,trueba2019comprehensive,kushwaha2023ixpe,rawat2023spectropolarimetric,zhang2026evidence}.

Particularly relevant here is its heartbeat-like variability, first reported during the 1998 outburst and seen again in 2021, when the source evolved from a hard state showing a low-frequency type-C QPO to a lower-frequency heartbeat state \citep{trudolyubov2001rxte,yang2022insight,zhao2023mhz,chen2025quasiperiodic,fan2025nicer}. 
Previous studies have left several questions open. 
The origin of type-C QPOs remains debated, and it is still unclear whether their modulation is tied mainly to the disk, the Comptonized flow, or changes in disk--corona coupling (see \citealt{ingram2019review} for a review). 
The origin of millihertz-scale
 QRM (hereafter "mHz QRM") is also uncertain, since heartbeat-like thermal--viscous cycles, radiation-pressure instability, and other low-frequency accretion-flow modulations can produce similar phenomenology \citep{belloni1997unified,neilsen2011physics,neilsen2012radiation}. 
4U~1630$-$47 is a useful source for addressing these questions because it shows recurrent outbursts, complex state evolution, repeated LFQPOs and heartbeat-like variability, disk winds, reflection features, and recent evidence of soft-state outflow changes involving both winds and jets.
In 2025, 4U~1630$-$47 entered a new outburst phase, which began around 
modified Julian date
 (MJD) 60768 (early April 2025) and was subsequently monitored extensively \citep{2025ATel17164....1M,parra202520}.

We studied the 2025 outburst 
 of 4U~1630$-$47 using the Neutron Star Interior Composition Explorer (\textit{NICER}). We investigated  both the transient LFQPOs detected during the rising phase and the later mHz QRM, and examined how their spectral properties evolve on short timescales. To do this, we combined standard timing analysis with wavelet-based interval selection and phase-resolved HHT analysis. Sect.~\ref{data} describes the observations and data reduction. In Sect.~\ref{ana}, we present the timing and spectral results, including the outburst evolution, the wavelet-based QPO separation, the corresponding spectral analysis, and the phase-resolved study of the QRM. In Sect.~\ref{dis}, we discuss the implications of these results. Our conclusions are summarized in Sect.~\ref{conclusion}.

\section{Data}
\label{data}

\subsection{Observations}
We used  \textit{NICER} \citep{gendreau2016neutron}, an X-ray timing instrument deployed on the International Space Station (ISS) as a NASA Explorers Mission of Opportunity, to monitor the outburst evolution. \textit{NICER} was launched on 3 June 2017 and installed on the ISS on 13 June 2017, providing a dedicated facility for high-throughput, soft X-ray observations. Its array of silicon-drift detectors covers the $0.2$--$12\,\mathrm{keV}$ energy band with a large effective collecting area ($>2000\,\mathrm{cm}^2$ at $1.5\,\mathrm{keV}$ and $\sim 600\,\mathrm{cm}^2$ at $6\,\mathrm{keV}$). The instrument time tags individual photons with $\sim 300\,\mathrm{ns}$ precision, which makes it well suited for tracking rapid aperiodic variability and searching for QPOs in accreting compact objects. In 2025, \textit{NICER} carried out an extended monitoring campaign of 4U~1630$-$47, providing dense coverage over the course of the outburst evolution. The campaign spanned from 7 April 2025 to 7 June 2025 and consisted of repeated pointings with exposure times ranging from a few hundred seconds to several kiloseconds.

\subsection{Data reduction }
For the \textit{NICER} data reduction, we followed the official \textit{NICER} analysis threads\footnote{\url{https://heasarc.gsfc.nasa.gov/docs/nicer/analysis_threads}}, using HEASoft~v6.36 and the \textit{NICER} calibration files (xti20240206). We applied standard calibration and screening with \texttt{nicerl2}. We extracted 1--10~keV light curves with \texttt{nicerl3-lc} and generated source/background spectra, together with the corresponding response files, using \texttt{nicerl3-spect}. Background estimation was performed with the \textsc{SCORPEON} background model.

We computed PDS with the \texttt{powerspec} task. The light curves were binned at $0.0078125\,\mathrm{s}$, corresponding to a Nyquist frequency of $64\,\mathrm{Hz}$. To search for possible QRM signals we divided each light curve into $256\,\mathrm{s}$ segments, yielding a minimum sampled frequency of $\simeq 3.9\times 10^{-3}\,\mathrm{Hz}$, adequate for studying heartbeat variability at a few millihertz. To search for QPO signals we used $64\,\mathrm{s}$ segments; this choice was motivated by the limited exposure of several observations. The PDS computed for individual segments were averaged for each observation and then geometrically rebinned in frequency with a factor of 1.05. Finally, we normalized the PDS to units of $\mathrm{rms}^2\,\mathrm{Hz}^{-1}$ and subtracted the Poisson noise \citep{belloni1990atlas,miyamoto1991x}.

We modeled the PDS in XSPEC v12.15.1 \citep{arnaud1996xspec} using a combination of Lorentzian components to describe both the BBN continuum and the QPO features. The fractional rms amplitude of the QPO was computed as the square root of the integrated power of the corresponding Lorentzian.  We estimated the uncertainties of the parameters using Markov Chain Monte Carlo (MCMC) simulations with 200 walkers and a length of $10^{5}$ steps. Parameter uncertainties correspond to the 90 percent credible interval derived from the posterior distributions obtained via MCMC sampling. In addition, we divided the light curves into two energy bands (2$-$5~keV and 5$-$10~keV) and computed the hardness ratio as the count-rate ratio between the hard and soft bands. 
\section{Analysis and results}
\label{ana}
\subsection{Light curve and PDS}
\begin{table*}
\centering
\caption{\textit{NICER} observations of 4U~1630$-$47. }
\label{tab:nicer}
\footnotesize
\setlength{\tabcolsep}{8pt}
\renewcommand{\arraystretch}{0.78}
\begin{tabular}{cccc|cccc}
\hline\hline
ObsID & Time & Exposure & PDS & ObsID & Time & Exposure & PDS\\
& MJD & (s) & & & MJD & (s) &\\
\hline
8130010105 & 60772.88 & 1556.99 & -- & 8130010120 & 60787.03 & 2400.00 & --\\
8130010106 & 60773.02 & 951.00 & QPO & 7717010303 & 60787.87 & 1247.00 & --\\
8130010107 & 60774.24 & 565.00 & QPO & 7717010304 & 60788.00 & 1363.00 & --\\
8130010108\tablefootmark{a} & 60775.21 & 1694.00 & -- & 8130010121 & 60788.32 & 488.00 & --\\
8130010109 & 60776.44 & 895.00 & QPO & 8130010122 & 60789.14 & 1273.00 & --\\
8130010110 & 60777.16 & 1446.00 & QPO & 8130010123 & 60791.21 & 770.00 & --\\
8130010111 & 60777.99 & 953.00 & QPO & 7717010401 & 60791.66 & 1884.00 & --\\
7717010101 & 60778.49 & 2267.00 & QPO & 7717010402 & 60792.05 & 191.00 & --\\
7717010102 & 60779.09 & 260.00 & -- & 8130010124 & 60792.96 & 99.00 & --\\
8130010112 & 60779.41 & 1499.00 & -- & 8130010125 & 60794.37 & 735.00 & --\\
8130010113 & 60779.99 & 7547.00 & QRM & 8130010126 & 60795.14 & 2739.42 & --\\
8130010114 & 60781.09 & 4468.00 & -- & 7721010201 & 60796.63 & 1771.00 & --\\
8130010115 & 60782.06 & 2600.00 & -- & 8130010127 & 60808.85 & 1257.00 & --\\
7721010101 & 60782.90 & 1740.00 & -- & 7717010305 & 60809.11 & 3768.00 & --\\
7721010102 & 60783.16 & 1575.00 & -- & 8531010101 & 60810.14 & 3892.00 & --\\
8130010116 & 60783.35 & 1435.00 & -- & 7717010403 & 60814.27 & 4103.00 & --\\
8130010117 & 60784.39 & 259.00 & -- & 8531010102 & 60816.08 & 3091.00 & --\\
7717010301 & 60784.64 & 2039.00 & -- & 8130010128 & 60817.44 & 2046.00 & --\\
7717010302 & 60785.03 & 1635.00 & -- & 8130010129 & 60820.09 & 809.00 & --\\
8130010118 & 60785.22 & 5830.00 & -- & 8531010103 & 60830.27 & 3171.00 & QPO\\
8130010119 & 60786.04 & 2221.00 & -- & 8130010130 & 60833.32 & 2382.00 & --\\
\hline
\end{tabular}
\tablefoot{ For each pointing, we list the ObsID, MJD, net exposure, and the corresponding PDS classification, where "--" indicates the absence of a prominent feature. \tablefoottext{a}{This observation shows a weak excess near $\sim0.31$~Hz, but its Lorentzian width is poorly constrained in both the best-fitting model and the MCMC posterior. It is therefore excluded from the subsequent analysis.}}
\end{table*}

\begin{figure}[t]
  \centering
  \includegraphics[width=0.85\columnwidth]{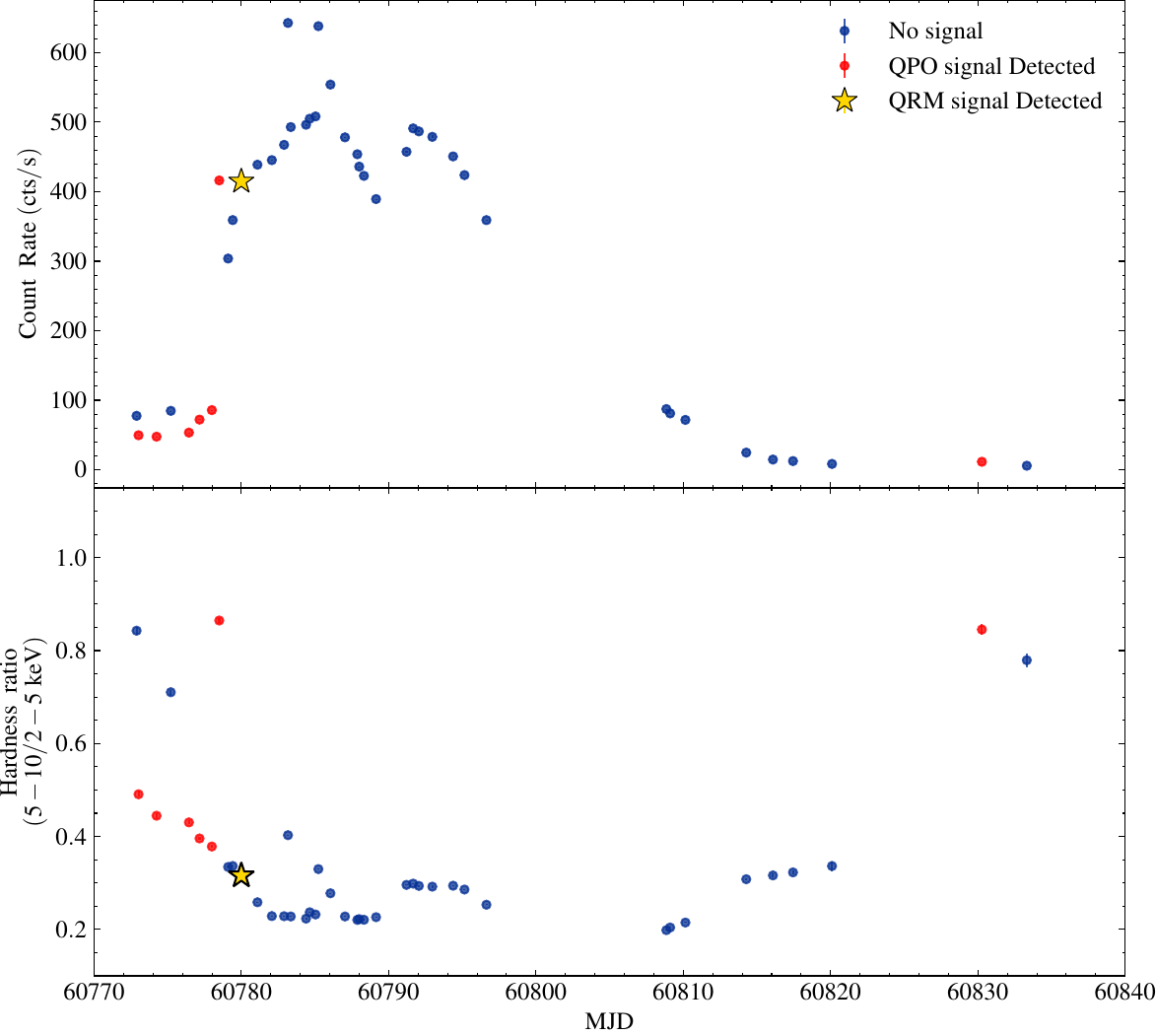}
  \caption{\textit{Top}: NICER light curve of 4U 1630$-$47 during the observations. Red circles mark the observations for which a QPO is detected in the PDS, and yellow stars indicate detections of a QRM. \textit{Bottom}: Corresponding hardness ratios; the symbols are the same as in the top panel.}
  \label{fig:fig1}
\end{figure}

\begin{figure}[t]
  \centering
  \includegraphics[width=0.85\columnwidth]{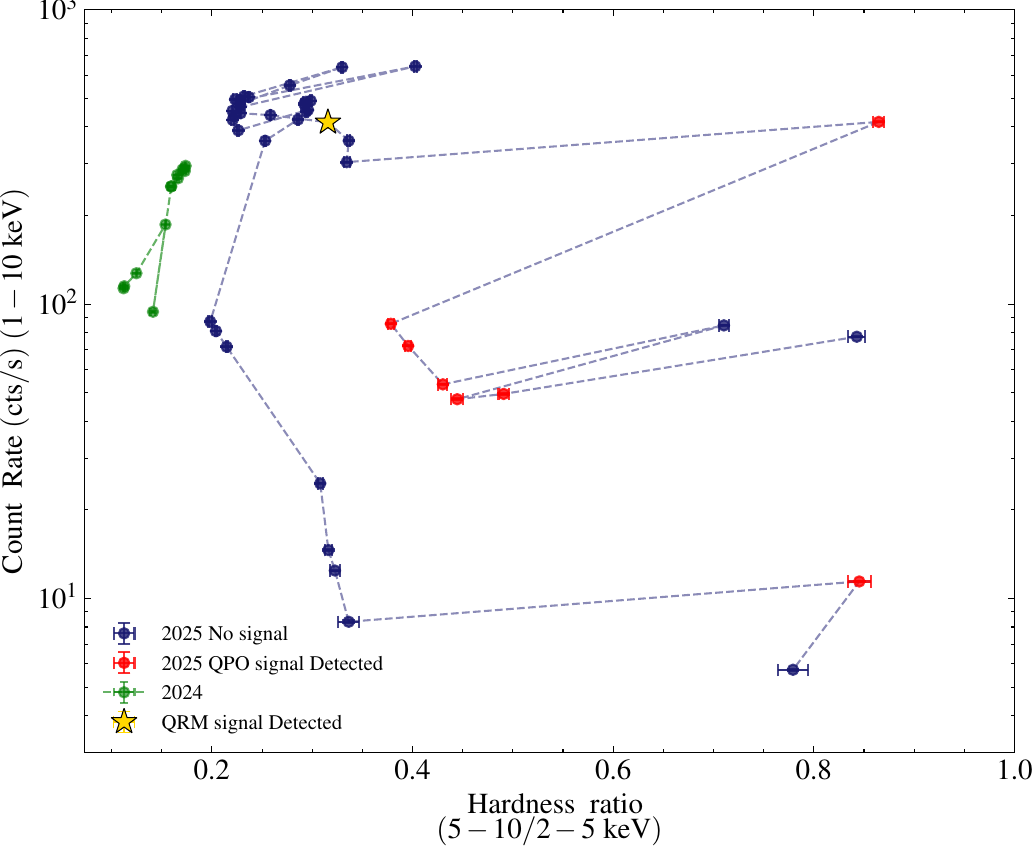}
  \caption{HID of 4U~1630$-$47. For comparison, we also plot the 2024 observations, marked with green points. The other symbols follow the same convention as in Fig.~\ref{fig:fig1}.}
  \label{fig:fig2}
\end{figure}

We summarize the \textit{NICER} observations of 4U~1630$-$47 in Table~\ref{tab:nicer}, listing the ObsID, observation time (MJD), exposure, and the presence of QPO/QRM features in the PDS.
We show the light curves and the corresponding hardness ratios for each observation in Fig.~\ref{fig:fig1}. 
Compared with the 2024 outburst, the 2025 observations provide a relatively complete coverage of the source evolution:
 the count rate is initially low and accompanied by a LFQPO (indicated by the red points in the figure), then rises rapidly to a peak of $\sim 600\,\mathrm{cts\,s^{-1}}$, and subsequently declines gradually. The hardness ratio is initially high, then decreases and approaches an approximately steady level, and finally increases again toward the end of the observations. The HID (Fig.~\ref{fig:fig2}) traces a relatively regular track: the source gradually softens from the hard state on the right, evolves along the left branch, and finally returns to the hard state in the bottom-right part of the diagram. For comparison, we also show the 2024 observations in the HID. 
These observations are mostly confined to the upper-left region of the HID and therefore do not trace the full outburst evolution. 
In addition, \cite{fan2025nicer} used the 2024 observations as part of their long-term spectral survey and to identify disk-wind features; no clear timing signal relevant to the present analysis was reported in those data. 
We therefore used the better-sampled 2025 outburst for the main timing and spectral analysis.

\begin{figure*}[t]
  \centering
  \includegraphics[width=0.8\columnwidth]{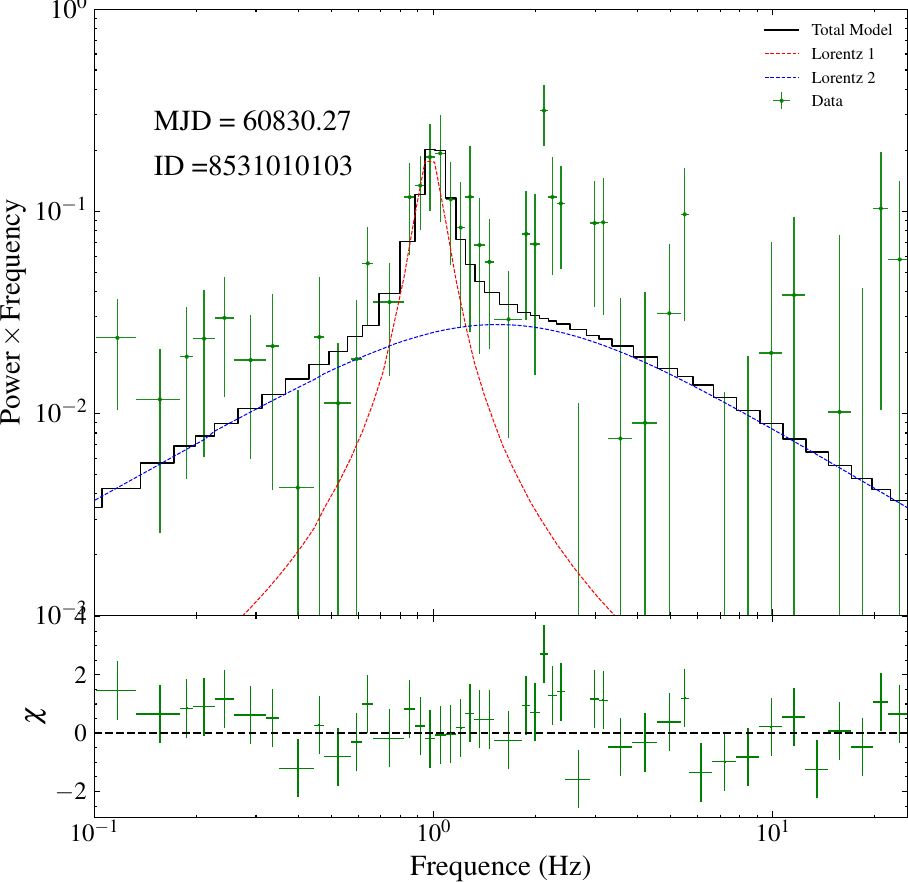}\includegraphics[width=0.8\columnwidth]{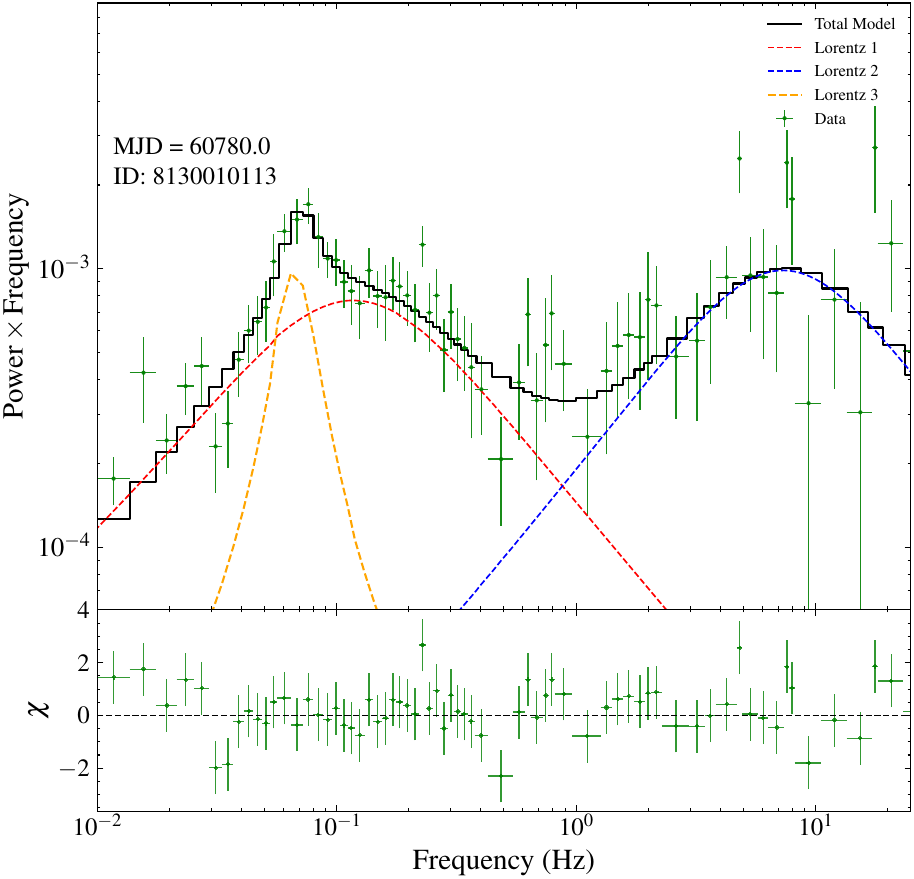}
  \caption{Representative fits to the PDS showing the QPO and QRM components in 4U~1630$-$47. \textit{Left}: Fit to the QPO signal. Green points show the data, the red curve the QPO component, the blue curve the BBN, and the black curve the total model; the lower subpanel shows the residuals defined as $\rm (data-model)/error$. \textit{Right}: Fit to the QRM signal. The orange curve indicates the QRM component. The observation time and ObsID are labeled in each panel. 
  For display purposes, we re-binned the data in the plot.  
  }
  \label{fig:pds}
\end{figure*}

We  report all best-fitting QPO and QRM parameters in Table~\ref{tab:qpo_params}. We show representative fits in Fig.~\ref{fig:pds}, and the QPO centroid frequency increases as the count rate rises, and no significant signal is detected after the frequency reaches $\sim 3.4\,\mathrm{Hz}$, until a $\sim 1\,\mathrm{Hz}$ QPO reappears toward the end of the outburst. The detected QPOs are classified as type-C QPOs. 
This classification is based on their centroid frequencies, relatively strong fractional rms amplitudes, association with BBN, and occurrence during the hard and intermediate part of the outburst evolution. 
These properties are more consistent with type-C QPOs than with type-B QPOs \citep{casella2005abc}.
We discuss QRM in more detail in Sect.~\ref{sec:qrm}.

\begin{table}[t]
  \centering
  \caption{Best-fitting QPO and QRM parameters.}
  \label{tab:qpo_params}
  \footnotesize
  \setlength{\tabcolsep}{3pt}
  \renewcommand{\arraystretch}{1.15}
  \begin{tabular}{lcccc}
    \hline\hline
    ObsID & $\nu$ (Hz) & FWHM (Hz) & rms (\%) & $\chi^{2}$/d.o.f.\\
    \hline
    8130010106 & $0.24\pm0.01$ & $0.09\pm0.03$ & $12_{-3}^{+2}$ & 89/88\\
    8130010107 & $0.29\pm0.02$ & $0.12_{-0.05}^{+0.02}$ & $14_{-3}^{+2}$ & 120/88\\
    8130010109 & $0.41\pm0.01$ & $0.10\pm0.04$ & $14_{-2}^{+2}$ & 103/88\\
    8130010110 & $1.18\pm0.04$ & $0.12\pm0.04$ & $8_{-2}^{+2}$ & 102/88\\
    8130010111 & $2.20\pm0.06$ & $0.34\pm0.14$ & $11_{-2}^{+2}$ & 118/88\\
    7717010101 & $3.43\pm0.14$ & $1.02\pm0.42$ & $7_{-2}^{+1}$ & 102/88\\
    8130010113\tablefootmark{a} & $0.069\pm0.005$ & $0.028_{-0.012}^{+0.024}$ & $4.7_{-0.5}^{+0.3}$ & 144/122\\
    8531010103 & $0.99\pm0.08$ & $0.18_{-0.09}^{+0.04}$ & $24_{-9}^{+7}$ & 89/88\\
    \hline
  \end{tabular}
  \tablefoot{Rows list the ObsIDs for which a QPO or QRM is detected. Columns give the centroid frequency $\nu$, FWHM, fractional rms amplitude, and fit statistic. \tablefoottext{a}{Detected mHz signal, classified here as QRM.}}
\end{table}

\subsection{Wavelet analysis}
With the development of time--frequency analysis techniques, it is now possible to probe the variability properties of accreting sources in much greater detail. In this work we therefore employed a wavelet transform \citep{torrence1998practical} to investigate the time--frequency or spectral behavior of the QPOs detected during the 2025 outburst, and to assess whether these signals are persistent or intermittent within each \textit{NICER} exposure.

Wavelet analysis provides a natural complement to standard Fourier methods because it retains information on when (in time) power at a given characteristic frequency is present. The continuous wavelet transform (CWT) of a discrete time series ($x_n$) is defined as the convolution of $x_n$ with a scaled and translated wavelet function ($\Psi$):
\begin{equation}
    W_n(s)=\sum_{n'=0}^{N-1} x_{n'} \Psi^*\left[\frac{(n'-n) \delta t}{s}\right],
\end{equation}
where $s$ is the wavelet scale, $n$ is the time index, $\delta t$ is the sampling interval, and the asterisk denotes complex conjugation. We used the complex Morlet wavelet with a dimensionless frequency $\omega_0 = 6$ as the mother wavelet and performed the CWT with the \texttt{pycwt} package\footnote{\url{https://github.com/regeirk/pycwt}}. The wavelet power spectrum is defined as $P_n(s)=|W_n(s)|^2$. We adopted $\omega_0 = 6$, a standard choice in practical applications, because it provides a suitable balance between time and frequency resolution. The statistical significance of the wavelet power was evaluated against a univariate lag-1 autoregressive [AR(1)] red-noise background following \citet{torrence1998practical}.

\begin{figure}[h]
  \centering
  \includegraphics[width=0.5\textwidth]{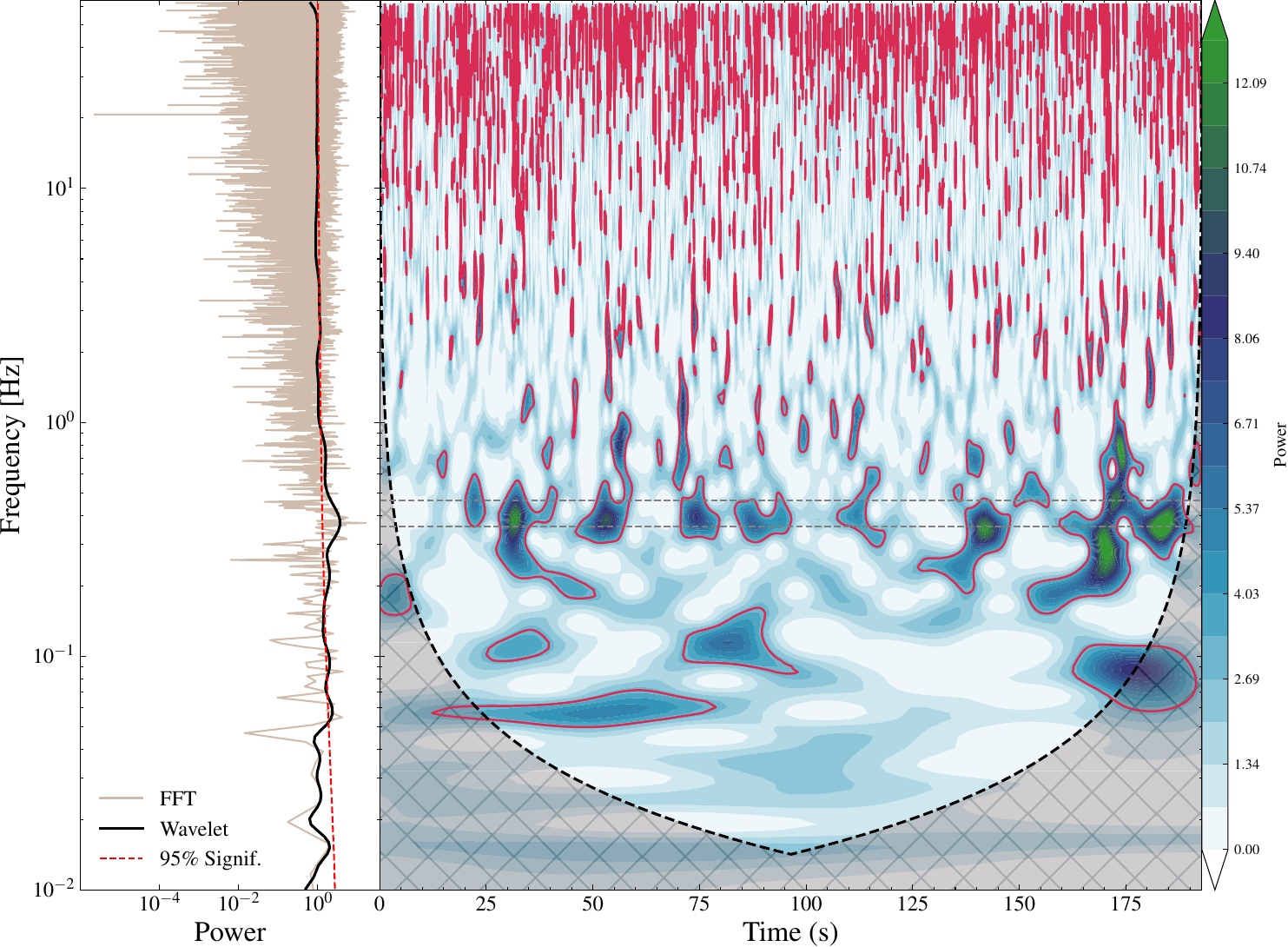}
  \caption{Wavelet analysis for the first GTI segment of ObsID 8130010109 of 4U~1630$-$47. \textit{Left}: Global wavelet power spectrum (black) compared with the time-averaged PDS (gray). The 95\% significance level is shown with a dashed red line.  \textit{Right}: Wavelet power spectrum as a function of time and frequency. The contour outlines regions exceeding the 95\% significance level, and the hatched area indicates the COI. The two horizontal gray lines indicate the QPO width obtained from the Lorentzian fit.}
  \label{wave}
  
\end{figure}

To minimize edge effects, we excluded the region inside the cone of influence (COI), where the wavelet power is affected by the finite length of the time series \citep{torrence1998practical}. Because smooth, continuous variations in the time series improve the robustness of the wavelet transform and because the \textit{NICER} light curves often contain gaps due to good time intervals (GTIs), we performed the wavelet analysis on each GTI segment separately. Once the peak frequency of the global wavelet power curve is confirmed, the 95 percent confidence interval of the frequency can be chosen, that is, the intersection range between the global wavelet power curve and the 95 percent confidence red line near the peak (Fig.~\ref{wave}). Then, if any point in the local wavelet plot within this frequency interval is greater than the 95 percent confidence level , the point is selected as the QPO time range \citep{chen2022waveleta,chen2022wavelet}. Using this method, we identified the QPO time intervals in each observation. We then used \texttt{ftcreate} to generate new GTI files for the selected intervals and ran\texttt{nicerl3-lc} and \texttt{nicerl3-spect} with these GTIs to obtain the corresponding light curves and spectra.

As an illustrative example, Fig.~\ref{wave} displays the wavelet-transform results for the first GTI segment of ObsID 8130010109. The left panel shows the global wavelet power spectrum overlaid with the corresponding time-averaged PDS and the 95\% significance level. The right panel presents the time--frequency wavelet power map.

\begin{figure}[h]
  \centering
  \includegraphics[width=0.4\textwidth]{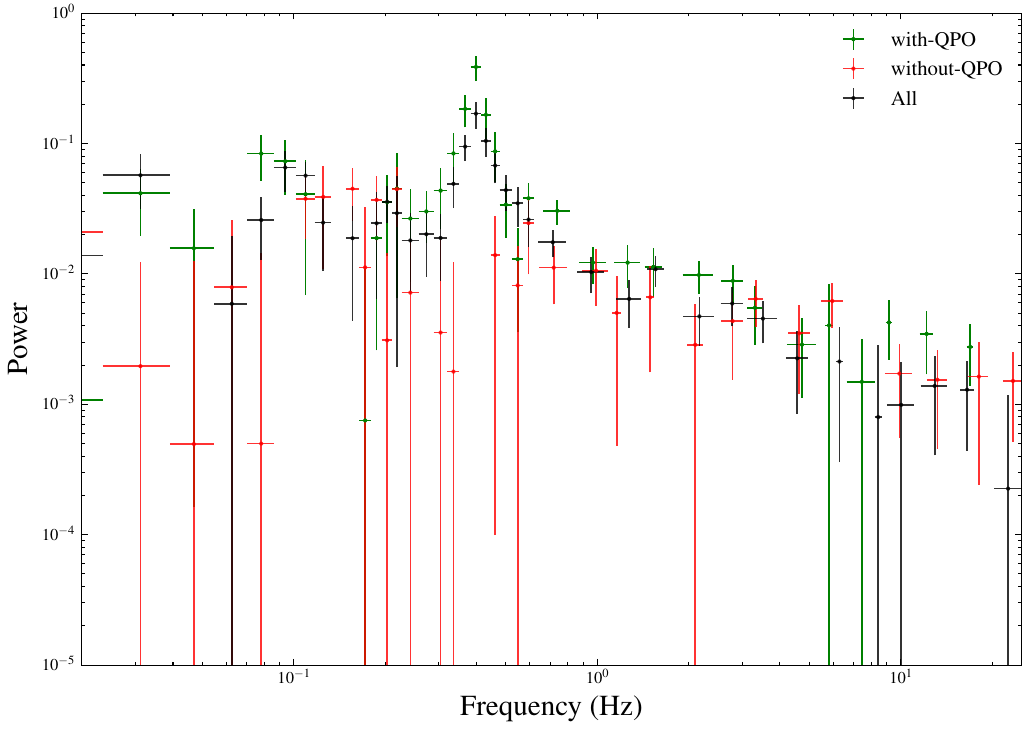}
  \caption{Comparison of PDS for ObsID 8130010109 of 4U~1630$-$47  derived from \textit{NICER} observations. The black points ("All") represent the average PDS of the entire observation. The green points ("with-QPO") show the averaged PDS computed from the light-curve segments identified by the wavelet transform as containing a QPO, while the red points ("without-QPO") are computed from the segments without a QPO signal. A distinct QPO feature is visible in the green data at $\sim 0.41,\mathrm{Hz}$, with a fractional rms amplitude of $\sim 17.5 \pm 0.02\%$, whereas no significant QPO is detected in the red data, for which the $3\sigma$ upper limit on the fractional rms amplitude is $< 5.36\%$, as described in the main text. 
  For display purposes, we rebinned the high-frequency data points.}
  \label{compare}
\end{figure}

Based on our wavelet analysis, we separated the light curves into segments with- and without-QPO signals. Figure~\ref{compare} presents the corresponding PDS for these selected segments using the same observation ID. It shows that the QPO feature is confined to a narrow frequency range around $\sim 0.41$ Hz, demonstrating that our method can effectively isolate the QPO-related component from the broadband aperiodic variability.

To further assess the effectiveness of the wavelet-based separation, we performed a constrained fit on the local PDS. By fixing the centroid frequency and FWHM at 0.41 Hz and 0.1 Hz, respectively, we derived a $3\sigma$ upper limit of $< 5.36\%$ for the fractional rms amplitude in the without-QPO segments. In contrast, the with-QPO segments exhibit a significant Lorentzian component with a fractional rms amplitude of $\sim 17.5 \%$. This large difference in power confirms that the wavelet-based selection successfully isolates the intervals where the transient QPO is physically present from those dominated by stochastic noise.

\subsection{Spectrum fitting} 

\begin{figure*}[htbp]
  \centering
  \includegraphics[width=0.55\columnwidth]{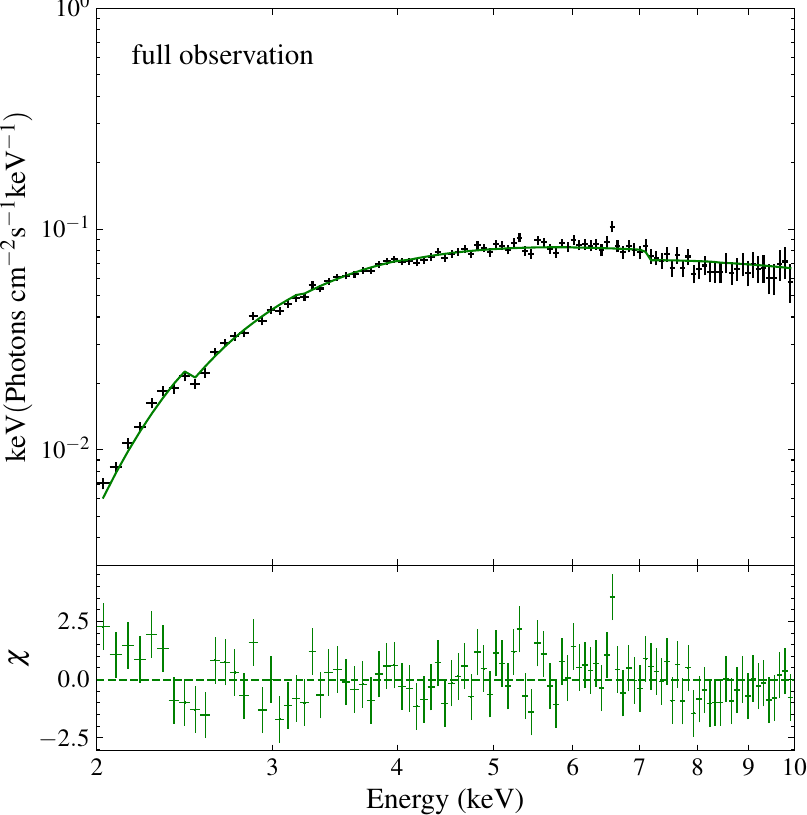}\includegraphics[width=0.55\columnwidth]{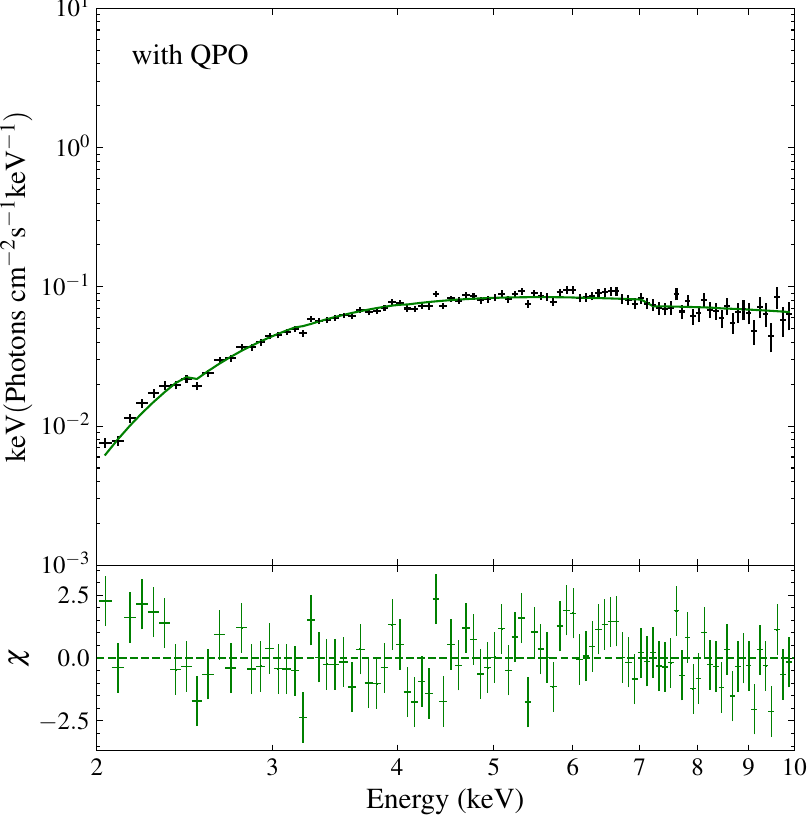}\includegraphics[width=0.55\columnwidth]{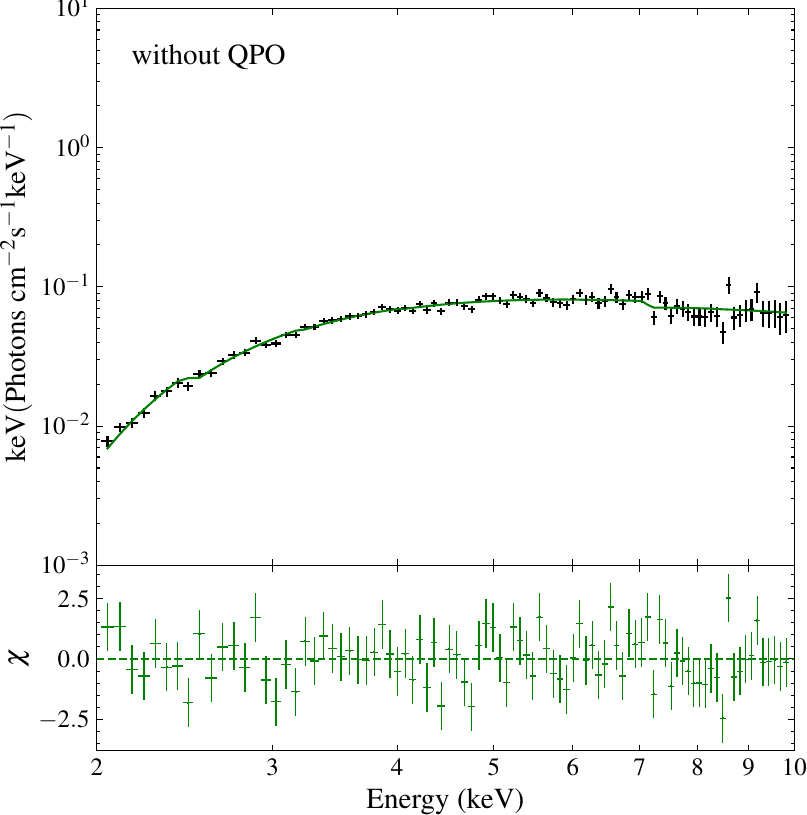}
  \includegraphics[width=0.55\columnwidth]{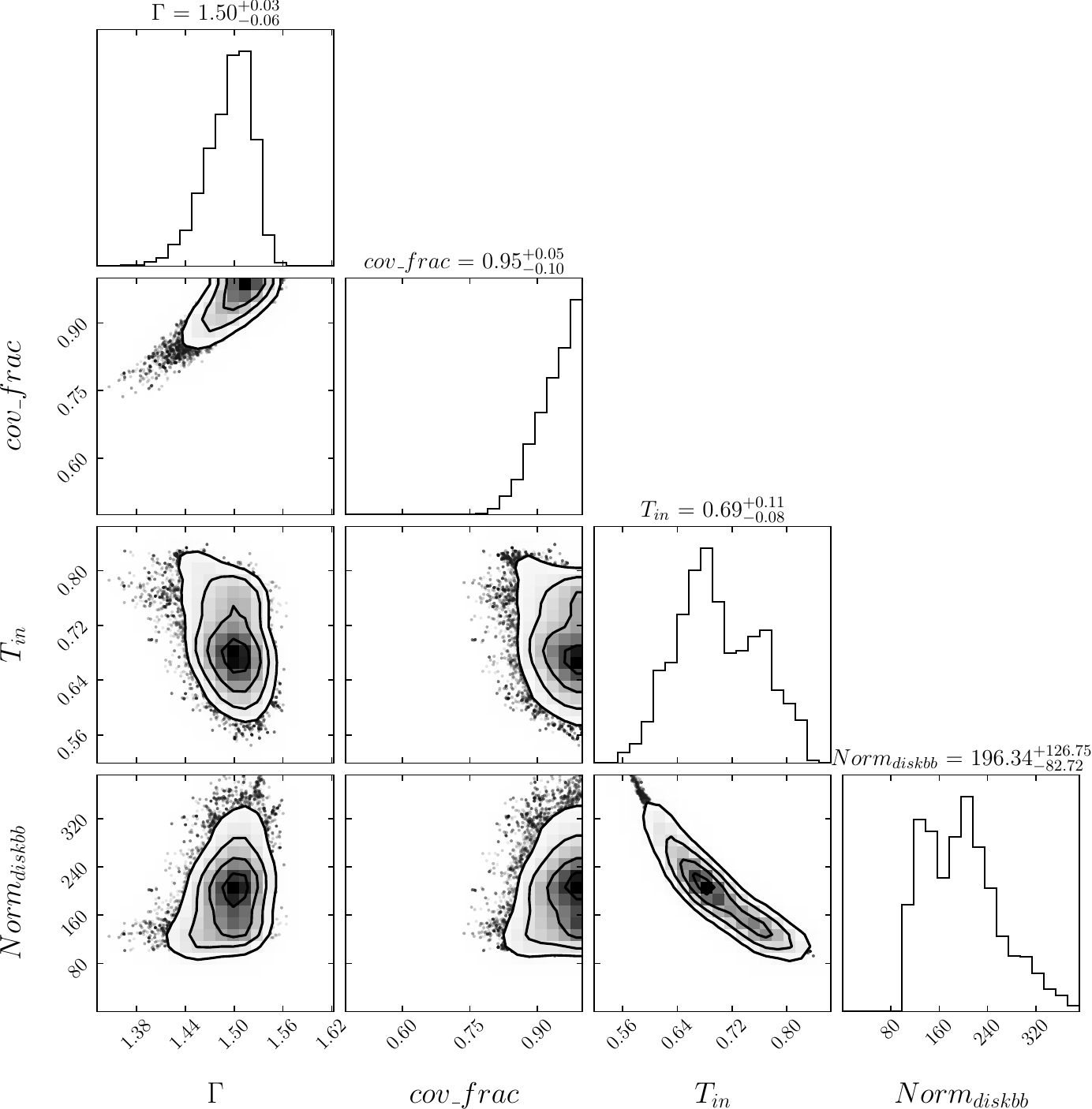}\includegraphics[width=0.55\columnwidth]{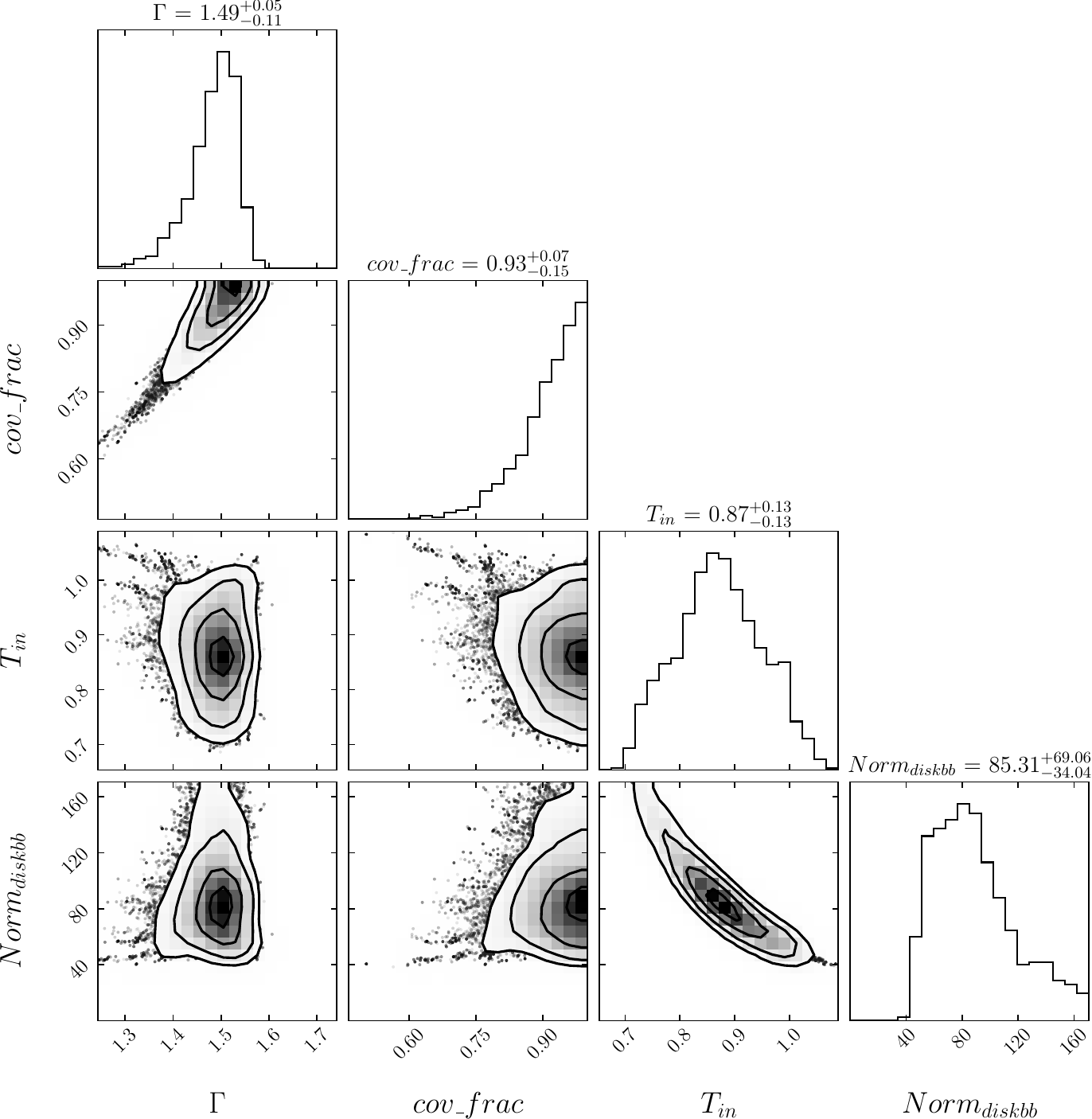}\includegraphics[width=0.55\columnwidth]{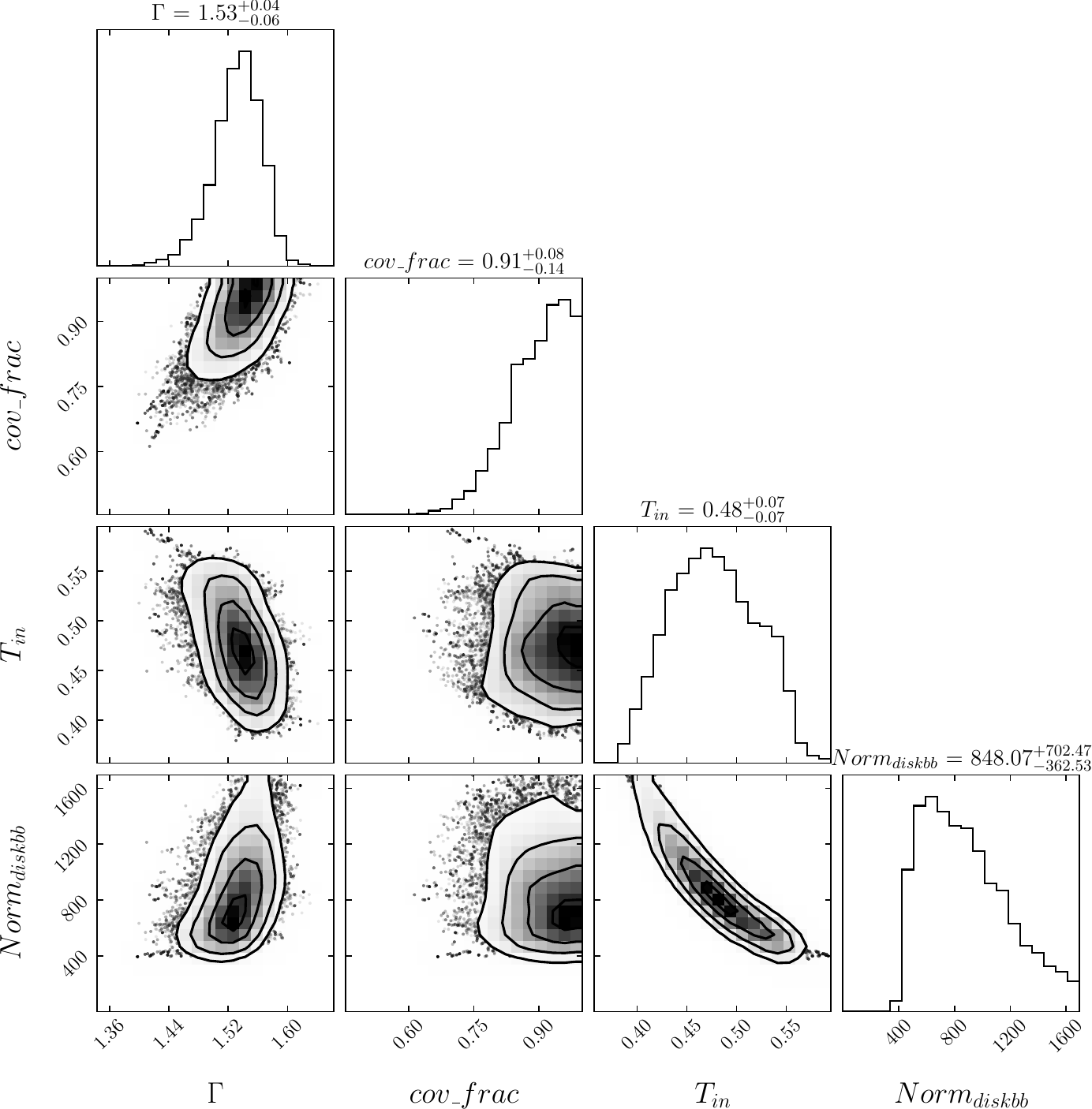}
  \caption{Spectral fitting results for observation 8130010109. \textit{Top row}: Energy spectra (black crosses) with the best-fit \texttt{tbfeo $\times$ (thcomp $\otimes$ diskbb)} models (solid green lines) in the 2--10 keV band, along with the fit residuals (lower subpanels). \textit{Bottom row}: Corresponding MCMC corner plots, illustrating the 1D posterior distributions and 2D confidence contours for the key parameters ($\Gamma$, $cov\_frac$, $T_{\rm in}$, and $\rm N_{\rm diskbb}$). Columns from \textit{left to right}: Spectra of the full observation, the time intervals with QPOs, and the time intervals without QPOs.}
  \label{fig:spec}
\end{figure*}

By employing wavelet analysis, we successfully separated the observations into time intervals with- and without-QPO signals. To investigate differences in the energy spectra between these two states, we extracted spectra using the \texttt{nicerl3-spect} pipeline with the corresponding GTIs.

Spectral fitting was performed using \texttt{PyXspec}\footnote{\url{https://heasarc.gsfc.nasa.gov/docs/software/xspec/python/html/index.html}}
, the Python interface to XSPEC v12.15.1 \citep{arnaud1996xspec}.  The spectral analysis was restricted to the 2-10 keV energy range. Due to strong interstellar absorption, the spectrum below 2 keV is heavily suppressed and largely dominated by the instrumental response shelf; therefore, this energy range was excluded from the fitting \citep{chopra2025spectro}.

To robustly estimate parameter uncertainties and explore potential degeneracies, we performed MCMC simulations using the Goodman-Weare algorithm. The MCMC chains were initialized with 32 walkers and run for a total of 48,000 steps. For each parameter, the median of the posterior distribution was adopted as the best-fit value. The errors on the fitted parameters are similarly derived from the  90 percent credible interval of the posterior distributions. 

Recently, \cite{fan2025nicer} presented a comprehensive spectral and timing analysis of \textit{NICER} observations of the BHXRB 4U 1630$-$47 spanning 2018-2024. To enable a direct comparison, we applied the same spectral model, \texttt{tbfeo $\times$ (thcomp $\otimes$ diskbb)}, to the energy spectra of the 2025 outburst.

The \texttt{tbfeo} component is an interstellar absorption model that allows the oxygen and iron abundances to be varied in addition to the hydrogen column density.We adopted the elemental abundances of \cite{wilms2000absorption} and cross-section takes of \cite{1996ApJ...465..487V}. When initially left free to vary, the absorption parameters were consistent with those reported by \cite{fan2025nicer}. To maintain consistency, we fixed the hydrogen column density and the oxygen and iron abundances to $N_{\rm H} = 16.96 \times 10^{22}~\mathrm{cm}^{-2}$, $A_{\rm O} = 0.21$, and $A_{\rm Fe} = 0.69$.

The thermal disk emission was modeled using the multicolor disk blackbody model \texttt{diskbb} \citep{mitsuda1985energy}, with free parameters given by the inner disk temperature ($kT_{\rm in}$) and the normalization ($N_{\rm diskbb}$). To describe the Comptonized component, we used \texttt{thcomp} \citep{zdziarski2020spectral}, a convolution model in which seed photons from the disk are inverse-Compton scattered by hot electrons in the corona. Its free parameters are the photon index ($\Gamma$), the electron temperature ($kT_{\rm e}$), the covering fraction ($cov_{frac}$), and the redshift.

We did not include ObsID~7717010101 in the subsequent analysis because the high-energy part of the 2--10~keV spectrum is strongly affected by the background. 
Although the source remains bright at  6--7 keV, the SCORPEON background becomes comparable to the  source spectrum above $\sim7$~keV, making the background-subtracted continuum poorly constrained. 
ObsID~8531010103 is also excluded from the spectral analysis, since the disk component is not significantly detected and its parameters cannot be robustly constrained in this observation. It is therefore not used in the subsequent comparison of $T_{\rm in}$ and $N_{\rm diskbb}$. 
Figure~\ref{fig:spec} shows the spectral fitting results and the corresponding MCMC posterior distributions for observation 8130010109, serving as a representative example.

\begin{figure}[h]
  \centering
  \includegraphics[width=0.4\textwidth]{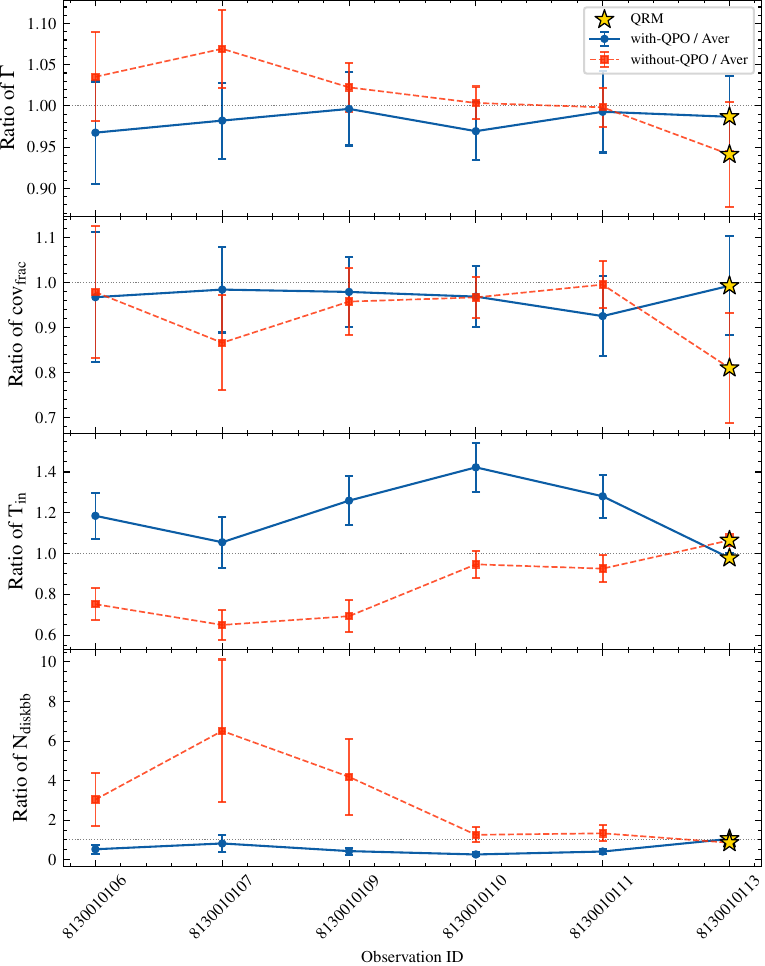}
  \caption{Evolution of the best-fit spectral parameters of 4U 1630$-$47 across different NICER observations. \textit{From top to bottom}: Ratios of the with-QPO/without-QPO best-fit parameters to their corresponding time-averaged values ($\Gamma$, $cov_{frac}$, and $\rm T_{\rm in}$) and $N_{\rm diskbb}$. The blue circles with solid lines represent the ratios of the parameters derived from the with-QPO intervals to those from the time-averaged (full) observations. The orange squares with dashed lines indicate the corresponding ratios for the without-QPO intervals. The star symbols mark the observation corresponding to the QRM. The horizontal dotted lines at 1.0 mark the reference level where the state-resolved parameter equals the time-averaged value. Error bars denote approximate 1$\sigma$ uncertainties on the ratios, propagated from the parameter uncertainties originally quoted at the 90\% confidence level. }
  \label{result1}
\end{figure}

\begin{table*}[t]
  \centering
  \caption{Spectral fitting parameters.  }
  \label{tab:fitparams}
  \setlength{\tabcolsep}{13pt}        
\renewcommand{\arraystretch}{1.25}
  
  \begin{tabular}{lccccc}
    \hline\hline
ObsID & $\rm T_{\rm in}$ (keV) & $N_{\rm diskbb}$ & $\Gamma$ & $ cov\_frac$  & $\chi^2$/d.o.f. \\
    \hline \multicolumn{6}{c}{Time-averaged} \\ \hline

8130010106 & $0.61_{-0.07}^{+0.09}$ & $286.92_{-115.33}^{+178.63}$ & $1.36_{-0.09}^{+0.07}$ & $0.77_{-0.12}^{+0.13}$  & 98.42/96 \\
8130010107 & $0.62\pm 0.09$ & $268.90_{-102.95}^{+229.27}$ & $1.48_{-0.08}^{+0.05}$ & $0.91_{-0.14}^{+0.09}$  & 74.54/90 \\
8130010109 & $0.69_{-0.08}^{+0.11}$ & $196.34_{-82.72}^{+126.75}$ & $1.50_{-0.06}^{+0.03}$ & $0.95_{-0.10}^{+0.05}$  & 91.88/96 \\
8130010110 & $0.63_{-0.06}^{+0.07}$ & $413.42_{-133.05}^{+212.29}$ & $1.63_{-0.05}^{+0.02}$ & $0.97_{-0.07}^{+0.03}$ & 111.19/100 \\
8130010111 & $0.77\pm 0.08$ & $245.52_{-75.35}^{+123.61}$ & $1.69_{-0.05}^{+0.03}$ & $0.97_{-0.08}^{+0.03}$  & 119.31/99 \\
8130010113 & $1.29_{-0.04}^{+0.05}$ & $197.42_{-25.44}^{+20.42}$ & $2.17_{-0.27}^{+0.11}$ & $0.86_{-0.25}^{+0.13}$ & 69.83/123 \\

\hline \multicolumn{6}{c}{QPO} \\ \hline
8130010106 & $0.72\pm 0.11$ & $151.56_{-61.09}^{+122.29}$ & $1.31_{-0.12}^{+0.11}$ & $0.74_{-0.14}^{+0.17}$  & 95.45/90 \\
8130010107 & $0.65_{-0.11}^{+0.14}$ & $220.80_{-113.05}^{+235.27}$ & $1.46_{-0.11}^{+0.07}$ & $0.89_{-0.17}^{+0.10}$  & 71.74/88 \\
8130010109 & $0.87_{-0.13}^{+0.13}$ & $85.31_{-34.04}^{+69.06}$ & $1.49_{-0.11}^{+0.06}$ & $0.93_{-0.15}^{+0.07}$ & 109.66/94 \\
8130010110 & $0.90_{-0.12}^{+0.14}$ & $109.41_{-44.25}^{+80.17}$ & $1.58_{-0.11}^{+0.06}$ & $0.94_{-0.16}^{+0.06}$  & 93.48/89 \\
8130010111 & $0.98_{-0.11}^{+0.13}$ & $101.26_{-37.36}^{+56.17}$ & $1.68_{-0.18}^{+0.09}$ & $0.90_{-0.23}^{+0.09}$  & 107.61/89 \\
8130010113 & $1.26\pm 0.03$ & $207.64_{-18.30}^{+19.99}$ & $2.14_{-0.13}^{+0.12}$ & $0.85_{-0.14}^{+0.13}$ & 68.48/120 \\

\hline \multicolumn{6}{c}{without-QPO} \\ \hline
8130010106 & $0.46_{-0.06}^{+0.08}$ & $872.56_{-406.63}^{+744.99}$ & $1.41_{-0.08}^{+0.07}$ & $0.75_{-0.16}^{+0.19}$  & 75.99/88 \\
8130010107 & $0.40_{-0.07}^{+0.08}$ & $1751.37_{-969.91}^{+1761.20}$ & $1.58_{-0.10}^{+0.08}$ & $0.78_{-0.21}^{+0.19}$  & 90.65/80 \\
8130010109 & $0.48\pm 0.07$ & $848.07_{-362.53}^{+702.47}$ & $1.54_{-0.06}^{+0.04}$ & $0.91_{-0.14}^{+0.08}$  & 91.42/87 \\
8130010110 & $0.60_{-0.05}^{+0.07}$ & $520.33_{-170.49}^{+234.92}$ & $1.64_{-0.06}^{+0.03}$ & $0.94_{-0.09}^{+0.06}$  & 96.92/98 \\
8130010111 & $0.71_{-0.08}^{+0.08}$ & $327.27_{-108.30}^{+177.56}$ & $1.69_{-0.06}^{+0.03}$ & $0.97_{-0.10}^{+0.03}$ & 96.90/96 \\
8130010113 & $1.37_{-0.05}^{+0.04}$ & $169.59_{-16.42}^{+24.03}$ & $2.04_{-0.23}^{+0.26}$ & $0.70_{-0.17}^{+0.25}$ & 83.37/113 \\
    \hline
  \end{tabular}
  \tablefoot{ The spectral model is \texttt{tbfeo $\times$ (thcomp $\otimes$ diskbb)}. A systematic error of 1.5\% was adopted when calculating $\chi^2$.}

\end{table*}

Figure~\ref{result1} illustrates the evolution of the relative spectral parameters across the analyzed NICER observations. To systematically compare the accretion states, we present the ratios of the best-fit parameters derived from the with-QPO (or with-QRM) and without-QPO (or without-QRM) intervals relative to their corresponding time-averaged, full-exposure values. Throughout the observations, both the $\Gamma$ and the covering fraction remain relatively stable. The without-QPO intervals consistently show a slightly softer power-law slope than the with-QPO intervals, although the discrepancy is modest and generally within the uncertainties.

In contrast, the accretion disk parameters differ between the with-QPO and without-QPO intervals. During the with-QPO intervals, $T_{\rm in}$ is higher by a factor of $\sim1.1$--$1.4$, and the without-QPO intervals are associated with a cooler disk. Meanwhile, $N_{\rm diskbb}$ is anticorrelated with temperature, taking lower values with-QPO intervals and higher values during the without-QPO intervals. As the outburst evolves to later epochs, the spectral differences between the two states gradually become smaller.  

Intriguingly, this evolutionary trend culminates in a complete phenomenological reversal during the final observation (ObsID 8130010113). Coinciding with the emergence of a QRM signal in the timing analysis, the wavelet spectrum spectral behavior is   inverted. Specifically, the inner disk temperature during the QRM intervals drops below that of the non-QRM intervals. Concurrently, the disk normalization reverses its previous pattern, showing a relative enhancement during the QRM intervals. The behavior of the photon index also flips accordingly. Collectively, this  inversion suggests a substantial structural shift in the accretion geometry during this final epoch. 

\subsection{Quasi-regular modulation}
\label{sec:qrm}
Hereafter, we use "mHz QRM" as the broader phenomenological term and reserve "heartbeat state" for the stronger 2023 event with a clear recurrent heartbeat-like morphology. 
The 2025 event is therefore described as mHz QRM rather than as a heartbeat state, because its rms amplitude is lower and its light curve is less regular.

\begin{figure}[htbp]
  \centering
  \includegraphics[width=0.5\textwidth]{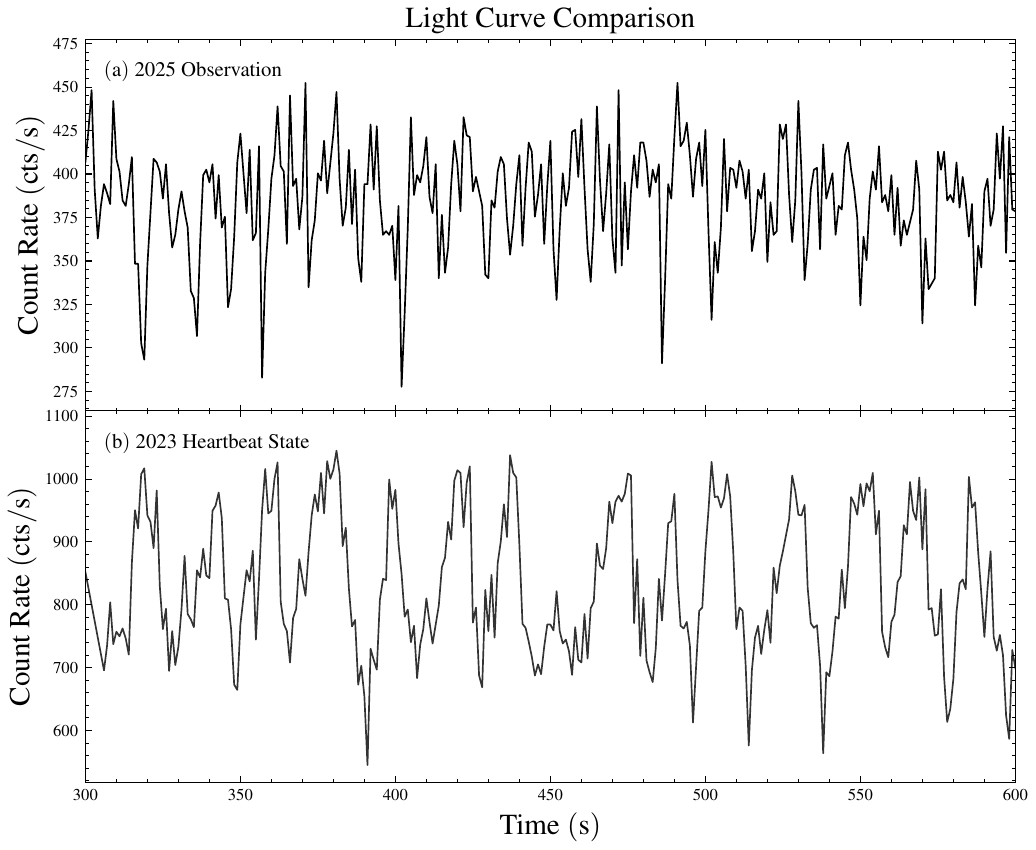}
  \caption{Comparison of the light curves of the QRM observed in 2025 (\textit{top}) and 2023 (\textit{bottom}; heartbeat state) of 4U 1630$-$47. The panels display 300-second segments of the NICER count rates.}
  \label{light}
\end{figure}

The QRM signal detected during the 2025 outburst has a centroid frequency of $\nu_{\rm QRM}\simeq 0.069$~Hz and  ${\rm FWHM}\simeq0.028$~Hz, corresponding to a quality factor of $Q\simeq2.5$. 
Its frequency range and $Q$ are consistent with those reported in earlier observations \citep{yang2022insight,chen2025quasiperiodic}, but its fractional rms amplitude is markedly lower, at $\sim4.7\%$. 
In Figure \ref{light}, we present a visual comparison between the 2025 data and the well-known  heartbeat state observed in 2023 of 4U 1630$-$47. During the 2023 observation, a QRM signal was detected at approximately 0.05 Hz with an rms amplitude of $\sim$10.6\%, manifesting as a highly pronounced modulation pattern. In contrast, the light curve profile of the QRM observed in 2025 is visibly less structured, reflecting its significantly suppressed rms amplitude. 

To investigate the physical evolution driving these distinct QRM cycles, we employed the HHT to obtain QRM phase-resolved spectra. This technique provides a robust, dynamic alternative to the traditional epoch-folding approaches utilized in previous studies \citep[e.g.,][]{neilsen2011physics, neilsen2012radiation, fan2025nicer}. The HHT analysis proceeds in two steps. First, we decomposed the light curve $x(t)$  into a sum of intrinsic mode functions (IMFs) and a residual trend $r_n(t)$ via complete ensemble empirical mode decomposition with adaptive noise (CEEMDAN; \citealt{torres2011complete}):
\begin{equation}
    x(t) = \sum_{j=1}^{n} c_j(t) + r_n(t),
    \label{eq:emd}
\end{equation}
\noindent where $c_j(t)$ represents the $j$-th IMF component. 

Subsequently, we isolated the dominant IMF corresponding to the QRM signal and applied the Hilbert transform,  $\mathcal{H}[\cdot]$, to construct its analytic signal, $z_j(t)$:
\begin{equation}
    z_j(t) = c_j(t) + i \mathcal{H}[c_j(t)] = a_j(t) e^{i \phi_j(t)},
    \label{eq:analytic_signal}
\end{equation}
\noindent where $a_j(t)$ is the instantaneous amplitude and $\phi_j(t) = \arctan\left(\frac{\mathcal{H}[c_j(t)]}{c_j(t)}\right)$ is the instantaneous phase. The instantaneous frequency is then defined as the time derivative of this phase, $\omega_j(t) = d\phi_j(t)/dt$. Unlike windowed transforms, this approach allows us to track the signal's frequency and phase content as a continuous function of time \citep[for comprehensive details on the HHT framework and its application to BHXRBs, see][]{huang2008review, zhu2025timing}.

Using the extracted instantaneous phase $\phi_j(t)$, each X-ray photon can be assigned to a specific phase interval, in a way directly analogous to standard phase-folding but without the assumption of strict periodicity. Since the HHT requires continuous data, the analysis was performed separately for each GTI. The instantaneous phase was then interpolated to the exact arrival time of every photon, and an extra phase-tag column was added to the \textit{NICER} event file. Based on this phase information, we extracted phase-resolved spectra to trace the evolution of the accretion flow over the QRM cycle. To improve the signal-to-noise ratio of the QRM signal, photons from time intervals with weak oscillation amplitudes, defined here as below 30\% of the local mean amplitude, were discarded \citep{su2015characterizing}.

The remaining valid photons were assigned normalized phase values between 0 and 1 and grouped into 10 equally spaced phase bins ($\Delta \phi = 0.1$). Phase-resolved source spectra were then extracted from the customized event list using the HEASoft \texttt{extractor} tool\footnote{\url{https://heasarc.gsfc.nasa.gov/docs/software/lheasoft/help/extractor.html}}. We also used the time-averaged background spectrum together with the standard redistribution matrix file and ancillary response file generated by the \texttt{nicerl3-spect} pipeline for the full observation. For spectral fitting, the exposure assigned to each phase bin was determined from the actual phase selection so that the flux normalization reflects the effective live time of that bin. This procedure allowd us to follow the spectral evolution of the source while accounting for cycle-to-cycle variations in both amplitude and period. For comparison, the same analysis was also applied to the 2023 data.

\begin{figure*}[htbp]
  \centering

  \includegraphics[width=0.8\columnwidth]{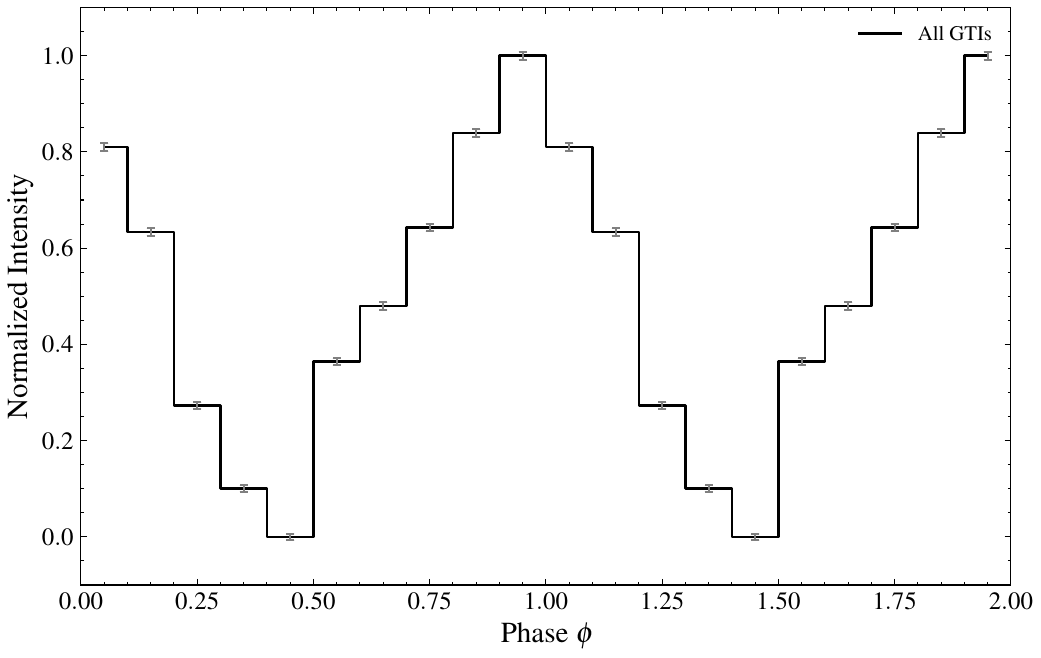}\includegraphics[width=0.8\columnwidth]{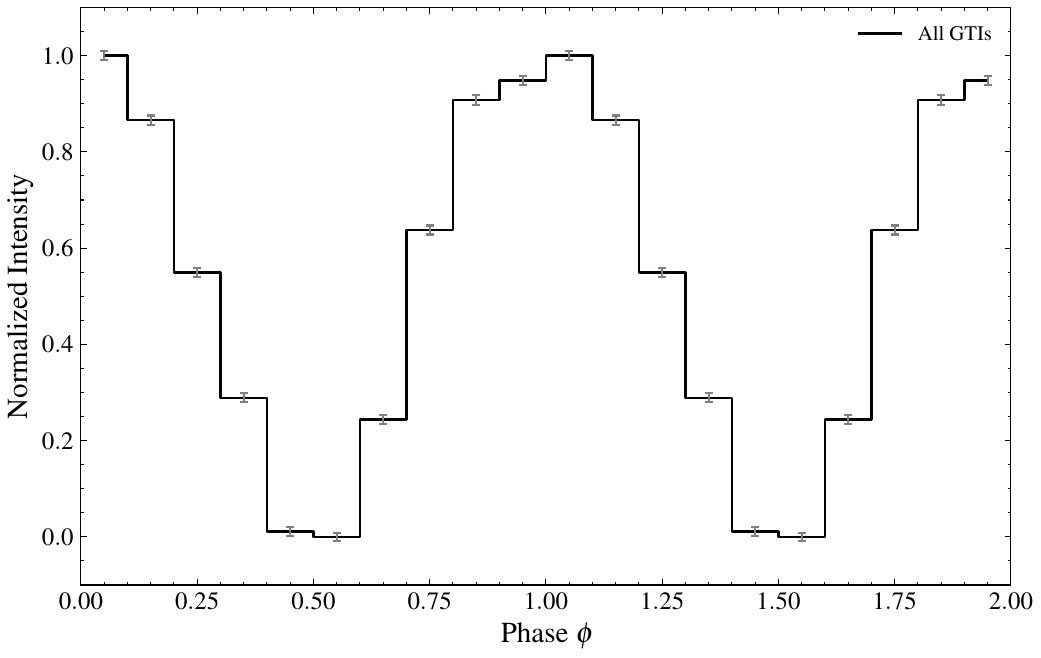}
  \includegraphics[width=0.8\columnwidth]{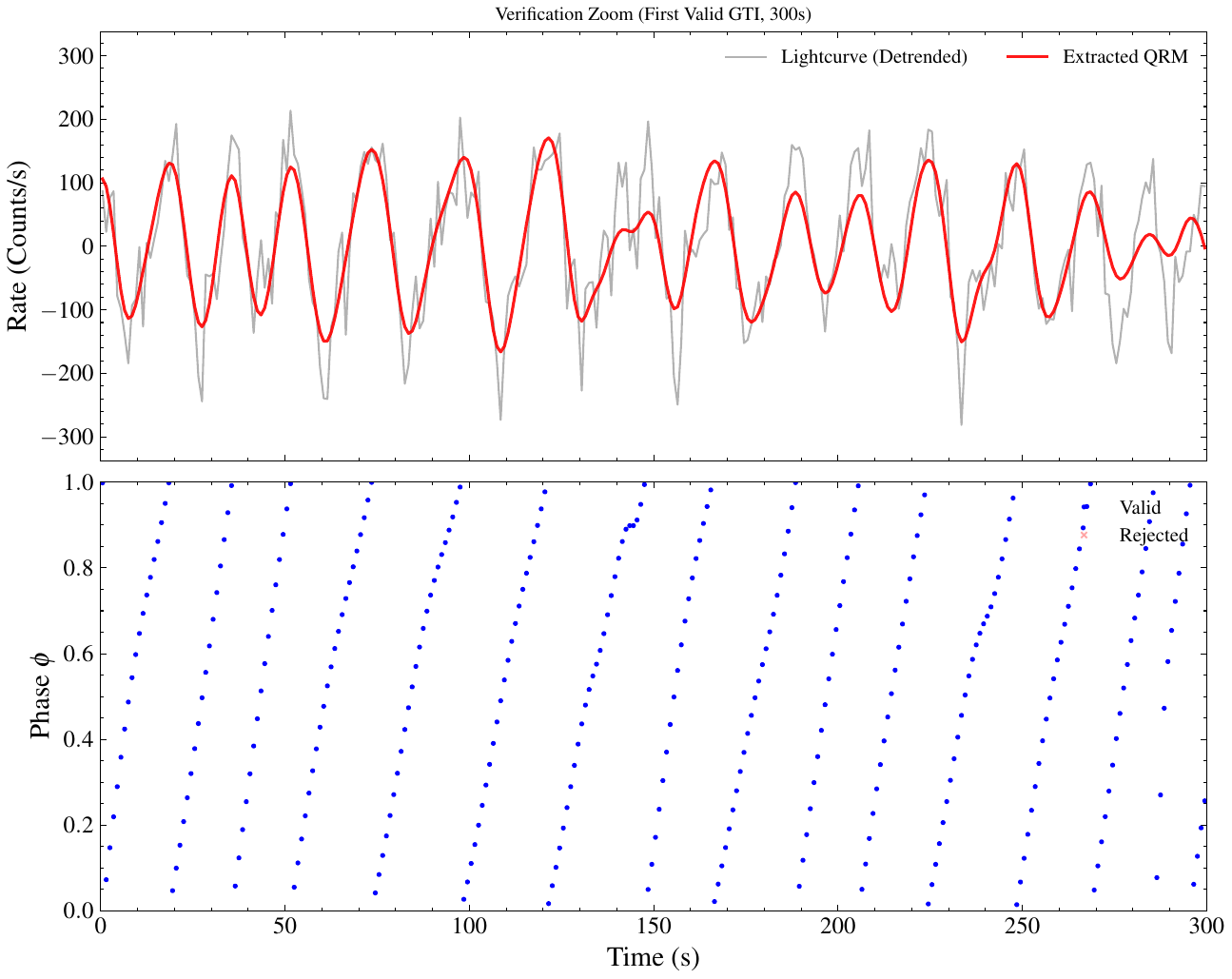}\includegraphics[width=0.8\columnwidth]{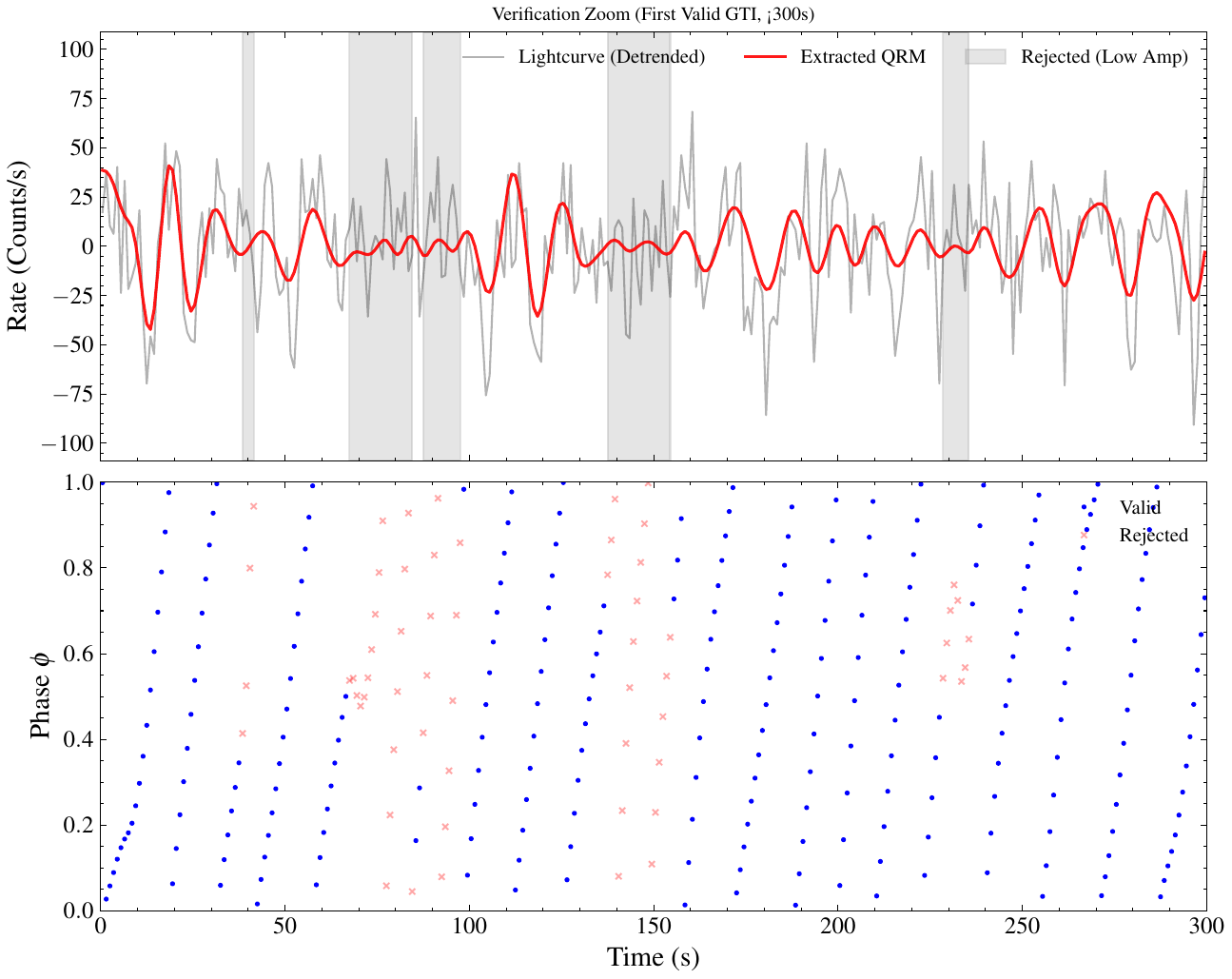}
  \caption{HHT phase-resolved and amplitude filtering for the 2023 (\textit{left}) and 2025 (\textit{right}) observations. \textit{Top}: Global phase-folded pulse profiles over two cycles. \textit{Middle}: 300-second segments of the detrended light curves (gray) and the extracted QRM signals (red). Shaded areas represent intervals rejected due to low instantaneous amplitude. For visual clarity and ease of comparison, the curves in the middle panels have been shifted to have a zero mean.   \textit{Bottom}: Instantaneous phase evolution. We show the valid phase bins used for the subsequent phase-resolved spectral extraction (blue dots)  and the discarded ones (red crosses).}
  \label{fig:hht_process}
\end{figure*}

To visually demonstrate the efficacy of our phase decomposition and filtering approach, we present a step-by-step visualization in Figure \ref{fig:hht_process}. Here, we compare the pronounced heartbeat state observed in 2023 (left column) with the much weaker QRM signal detected in our 2025 observation (right column). As illustrated in the middle panels, the CEEMDAN algorithm successfully tracks the dominant oscillatory component (solid red line) embedded within the light curve (gray line).

For the 2023 data, the modulation is sufficiently stable to allow continuous phase extraction. In contrast, the QRM observed in 2025 is much weaker (with a fractional rms of $\sim$4\%) and shows noticeable cycle-to-cycle amplitude variations. To reduce the influence of low-amplitude, noise-dominated intervals on the phase-resolved spectra, we applied the 30\% amplitude threshold described above. Intervals that did not satisfy this criterion were masked out dynamically. The selection range adopted here differs from that of \citet{su2015characterizing}, who defined the threshold from the average of the $3\sigma$ lower limit of the instantaneous amplitude; in the present work, we instead adopted a fixed quantitative criterion for simplicity. The bottom panels show the corresponding instantaneous phases, distinguishing between the valid phase tags retained and those rejected for spectral folding. As shown in the top panels, this selective phase-tagging procedure yields a cleaner folded pulse profile for both states.

\begin{figure*}[htbp]
  \centering
  \includegraphics[width=0.8\columnwidth]{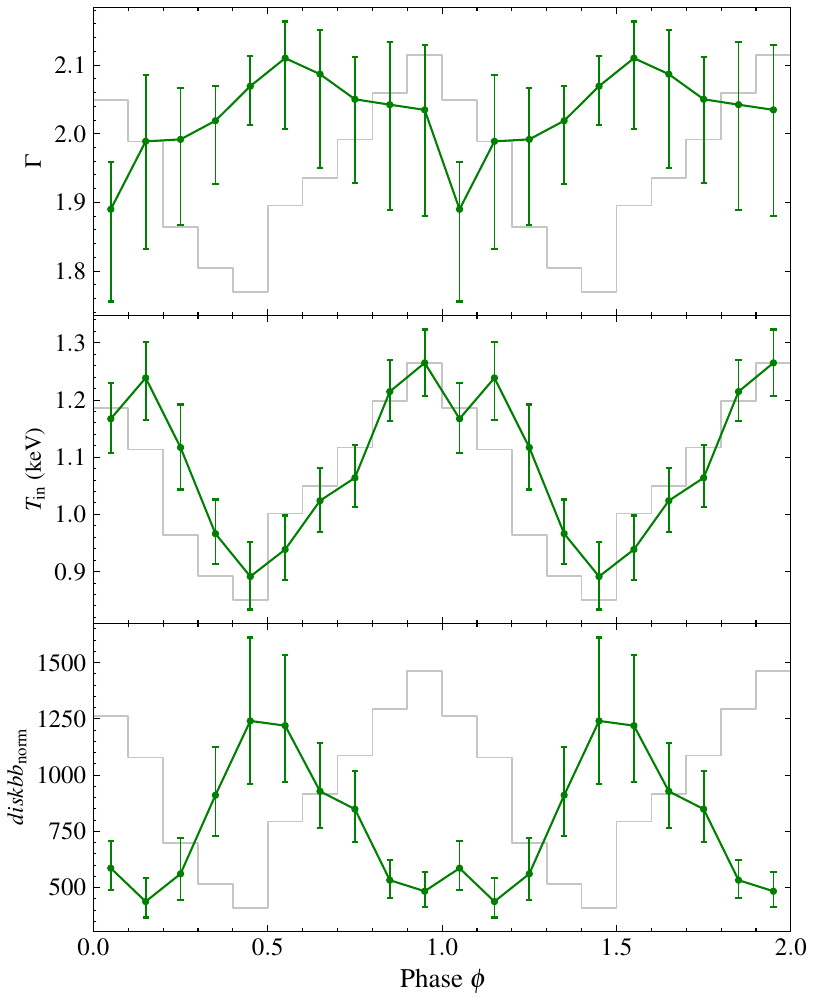}\includegraphics[width=0.8\columnwidth]{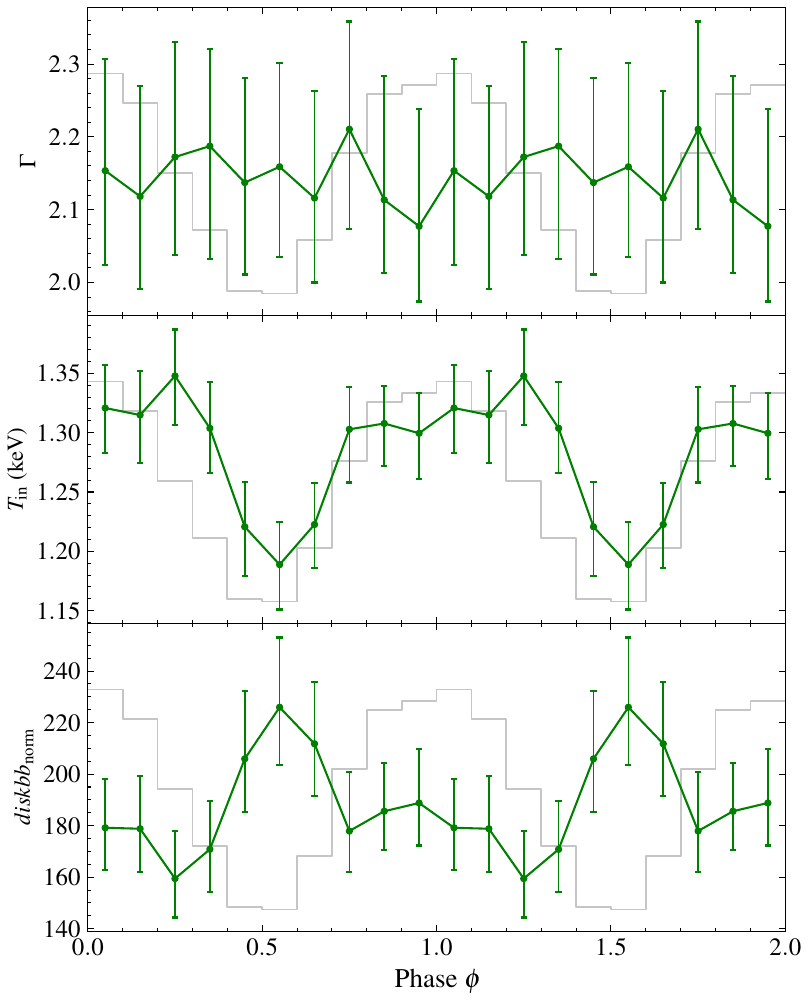}
  \caption{Phase-resolved spectral evolution for the  heartbeat state observed in 2023 (left column) and the QRM observed in 2025 (right column). From top to bottom, the panels display the phase dependence of $\Gamma$, $\rm T_{\rm in}$, and $N_{\rm diskbb}$, respectively. The green circles represent the best-fit parameter values derived from the MCMC posterior distributions, with error bars indicating the 90\% confidence intervals. For visual reference, the normalized, phase-folded light curve is overlaid in the background of each panel as a gray step function. All data are plotted over two continuous cycles ($0 \le \phi \le 2$).}
  \label{fig:hht_fit}
\end{figure*}

To characterize the detailed spectral evolution across the QRM cycle, we fitted each of the ten phase-resolved spectra using the same physical model adopted in the time-averaged analysis, \texttt{tbfeo $\times$ (thcomp $\otimes$ diskbb)}. During the fitting procedure, the interstellar absorption parameters ($N_{\rm H}$, as well as the oxygen and iron abundances) were fixed at their time-averaged values, while $\rm T_{\rm in}$, $N_{\rm diskbb}$, $\Gamma$,  and $cov_{frac}$ were allowed to vary freely with phase. The evolution of the spectral parameters over two consecutive cycles is presented in Figure~\ref{fig:hht_fit}, where the best-fit parameters are overlaid with the normalized pulse profiles (gray step curves) for reference. Since the covering fraction exhibits little variation over the entire phase, we do not show it here.

In the  heartbeat state observed in 2023, the spectral parameters exhibit highly significant modulations. The inner disk temperature  is in phase with the X-ray count rate, reaching its global maximum at the pulse peak ($\phi \approx 1.0$). In contrast, the disk normalization shows a pronounced anticorrelation with both the flux and $\rm T_{\rm in}$, attaining its minimum at the pulse maximum.

Remarkably, the phase-resolved spectral evolution of the QRM observed in 2025 follows the same physical pattern, although with a reduced modulation amplitude. The $T_{\rm in}$ profile remains tightly in phase with the pulse profile, while $\rm N_{\rm diskbb}$ preserves its systematic anticorrelation. The Comptonization photon index displays only marginal variations compared to the strong disk parameter modulations, indicating that the cyclic variability is primarily driven by changes in the accretion disk rather than the coronal emission.

To quantitatively assess the coupling between the spectral parameters and the observed X-ray count rates, we performed a statistical correlation analysis between the phase-resolved count rates and the corresponding best-fit spectral parameters. We adopted the Pearson correlation coefficient ($r$) to evaluate the linear dependence between the  variables\citep{benesty2009pearson}. For each correlation, we also computed the corresponding $p$-value to assess its statistical significance. Under the null hypothesis, the true correlation coefficient is zero ($H_0: \rho = 0$), whereas the alternative hypothesis is that the true correlation coefficient is nonzero ($H_1: \rho \neq 0$). We adopted 0.05 as the significance threshold, such that $p < 0.05$ indicates a statistically significant correlation. The resulting correlation coefficients and significance levels are summarized in Table~\ref{tab:relation}.

\begin{table}[t]
\centering
\caption{Correlation coefficients.}
\label{tab:relation}
\footnotesize
\setlength{\tabcolsep}{4pt} 
\renewcommand{\arraystretch}{1.4}
\begin{tabular}{ccccc}
\hline
\hline
\multirow{2}{*}{Parameters} & \multicolumn{2}{c}{2023} & \multicolumn{2}{c}{2025} \\
\cline{2-5}
                             &   $r$  & $p$ & $r$& $p$  \\
\hline
$\rm T_{\rm in}$                 & $0.82\pm0.19$ &$0.0038$&$0.88\pm0.15$&$0.0009$ \\
$\rm N_{\rm diskbb}$             & $-0.68\pm0.29$& $0.0333$ & $-0.73\pm0.32$ & $0.0204$ \\
$\Gamma$                     & $ -0.26\pm0.40$ & $0.4633$ & $-0.29\pm0.42$ & $0.4524$ \\
\hline
\end{tabular}
\tablefoot{Pearson correlation coefficient ($r$) and $p$-values for the correlation between phase-resolved spectral parameters and  count rate for the 2023 and 2025 observations of 4U 1630$-$47.}
\end{table}

For the $\rm T_{\rm in}$, we find a strong and highly significant positive correlation in both epochs. In the heartbeat state observed in 2023, the correlation coefficient is $r = 0.825 \pm 0.188$ ($p = 0.0038$).\ In the 2025 observation, the coupling is even stronger, with $r = 0.884 \pm 0.152$ and a higher significance level ($p = 0.0009$). Conversely, the $\rm N_{\rm diskbb}$ exhibits a significant anticorrelation with the count rate in both observations. For the 2023 data, we obtain $r = -0.681 \pm 0.296$ ($p = 0.0333$), and a similar trend is maintained in 2025 with $r = -0.731 \pm 0.324$ ($p = 0.0204$). This systematic decrease in $N_{\rm diskbb}$ during flux peaks suggests that the apparent disk radius effectively shrinks as the temperature increases, a behavior consistent with the radiation-pressure instability model. In contrast, $\Gamma$ shows no statistically significant correlation with the count rate in either observation, with $p$-values consistently exceeding $0.45$. The lack of a clear trend in $\Gamma$ implies that the physical properties of the corona remain relatively stable throughout the oscillation cycles, further reinforcing the conclusion that the observed variability is disk-dominated. 
\section{Discussion}

\label{dis}

\subsection{Intermittent type-C QPOs and their spectral dependence}

Wavelet analysis reveals that the type-C QPO varies substantially in strength within a single NICER exposure. 
We therefore divided each observation into intervals with and without a significant QPO signal, selected according to the presence or absence of significant wavelet power near the QPO frequency. 
The spectra extracted from these intervals show a consistent trend: the with-QPO intervals have a higher fitted disk temperature and a lower \texttt{diskbb} normalization than the without-QPO intervals, whereas the photon index varies less systematically and is generally consistent within the uncertainties (see Fig.~\ref{result1}). 
Within the NICER band, the spectral differences associated with the intermittent QPO are therefore most clearly seen in the fitted disk parameters.

We fitted the 2--10~keV spectra with \texttt{tbfeo $\times$ (thcomp $\otimes$ diskbb)} to analyze all intervals uniformly and to remain consistent with previous NICER studies of 4U~1630--47 \citep{fan2025nicer}. 
NICER alone, however, does not constrain the high-energy rollover of the Comptonized continuum and has limited leverage on possible reflection components, leaving some degeneracy between the disk and Comptonized emission \citep{done2007modelling}. 
The fitted values of $T_{\rm in}$, $N_{\rm diskbb}$, $\Gamma$, and the covering fraction should therefore be treated as model-dependent spectral tracers \citep{merloni2000interpretation,done2007modelling}. 
Our conclusions are based on relative differences measured with the same model, not on the absolute physical values of individual parameters.

The lower $N_{\rm diskbb}$ in the with-QPO intervals should not be read directly as a smaller physical inner radius. 
In \texttt{diskbb}, $N_{\rm diskbb}=(r_{\rm in}/D_{10})^2\cos i$, where $r_{\rm in}$ is the apparent inner disk radius. 
The physical radius is commonly written as $R_{\rm in}=\xi f_{\rm col}^{2}r_{\rm in}$, with $f_{\rm col}$ the color-correction factor and $\xi$ the inner-boundary correction \citep{kubota1998evidence,shimura1995spectral}. 
For fixed $R_{\rm in}$, distance, inclination, and $\xi$, this gives $N_{\rm diskbb}\propto f_{\rm col}^{-4}$, so a larger $f_{\rm col}$ in a hotter disk can mimic a decrease in $N_{\rm diskbb}$. 
For ObsID~8130010109, for example, $T_{\rm in}$ increases from $0.48$~keV in the without-QPO interval to $0.87$~keV in the with-QPO interval. 
If $f_{\rm col}\propto T_{\rm in}^{1/4}$, this corresponds to 
$f_{\rm col,QPO}/f_{\rm col,nonQPO}\simeq(0.87/0.48)^{1/4}\simeq1.16$, and spectral hardening alone would predict 
$N_{\rm diskbb,QPO}/N_{\rm diskbb,nonQPO}\simeq1.16^{-4}\simeq0.55$. 
The observed ratio is much smaller, $85.31/848.07\simeq0.10$. 
Thus, spectral hardening contributes in the correct direction but cannot by itself explain the full decrease in $N_{\rm diskbb}$.
Partial obscuration by the inner Comptonizing flow, disk warping, or changes in projected area can modify the apparent disk-emitting region seen by NICER. 
In addition, a varying fraction of disk photons intercepted and Comptonized by the corona can reduce the directly observed thermal disk flux. 
These effects can change the fitted $N_{\rm diskbb}$ without requiring the disk inner edge to move by the same amount \citep{merloni2000interpretation,steiner2009simple}. 
We therefore interpret the $T_{\rm in}$--$N_{\rm diskbb}$ anticorrelation mainly as a change in the apparent disk emission associated with the QPO, rather than as a direct measurement of inner-radius motion.

Wavelet-selected intervals with and without QPOs have previously been used to study transient type-C QPOs in MAXI~J1535--571 \citep{chen2022waveleta}, which remained in the hard-intermediate state while the QPO appeared and disappeared on short timescales. In 4U~1630--47, the QPO is also intermittent during the rising phase. 
The associated spectral behavior differs, however. The Insight-HXMT results for MAXI~J1535--571 pointed primarily to changes in the hard Comptonized component, whereas our \textit{NICER} results show the clearest differences in the fitted disk parameters. Given the limited \textit{NICER} coverage above $\sim$10~keV, it is difficult to exclude the possibility that the hard Comptonized component also varies significantly in 4U~1630--47.

Swift~J1658.2--4242 provides another example of rapid changes in QPO detectability. 
\citet{xu2019broadband} reported a flux drop of $\sim$45\% within $\sim$40~s, accompanied by the sudden appearance of a 6--7~Hz QPO and a large increase in broadband fractional rms. 
The high-flux interval showed no significant QPO while the low-flux interval showed a clear detection.  
In 4U~1630--47, by contrast, the source remains in a QPO-producing regime throughout the rising phase, with the QPO power fluctuating rather than switching on from quiescence.

A comparison with the 2021 outburst provides a useful reference for the long-term evolution. 
\citet{chopra2025spectro} reported a positive $\nu_{\rm QPO}$--$\Gamma$ correlation during the rising phase and interpreted it in terms of coronal evolution. 
Our 2025 data show the same qualitative trend, with $\Gamma$ increasing as the QPO frequency rises (Tables~\ref{tab:qpo_params} and \ref{tab:fitparams}). 
As shown in Fig.~\ref{fig:gamma_compare} in Appendix~\ref{app:gamam}, the slopes are consistent within the uncertainties, but the 2025 points are systematically offset to lower $\Gamma$. 
We therefore used the 2021 result as a qualitative reference rather than as a strict quantitative benchmark.

The $\nu_{\rm QPO}$--$\Gamma$ relation should be interpreted as a secular trend within the QPO-producing hard and intermediate state, not as evidence of rapid transitions between distinct spectral states. 
As $\nu_{\rm QPO}$ increases during the rising phase, the fitted spectrum softens \citep{vignarca2003tracing,titarchuk2004spectral}. 
This behavior may reflect a gradual evolution of the inner accretion geometry, for example a more compact Comptonizing region, stronger cooling of the hot flow by disk photons, or inward motion of the characteristic radius associated with the type-C QPO \citep{kubota2024disc,ma2025testing}. 
This long-timescale evolution is distinct from the wavelet-selected QPO intermittency, which probes changes in QPO power within individual NICER exposures and shows its clearest spectral signature in the fitted disk parameters.

\subsection{The QRM observed in 2025}

Figure~\ref{result1} marks the observation in which the mHz QRM is detected. 
This point differs from the earlier LFQPO detections in the interval-resolved comparison. 
During the LFQPO phase, the oscillating intervals are generally hotter and have lower apparent $N_{\rm diskbb}$ than the non-oscillating intervals; the same trend is also seen relative to the time-averaged spectra. 
In ObsID~8130010113, where the mHz QRM is detected, the ordering is reversed: the oscillating intervals are cooler and have a larger apparent $N_{\rm diskbb}$. 
The mHz QRM therefore does not appear to be a simple low-frequency extension of the LFQPO intermittency.

We applied HHT-based phase-resolved spectroscopy to the QRM observed in 2025 and to the  heartbeat state observed in 2023. 
The two observations show similar phase-resolved changes in the disk parameters: $T_{\rm in}$ broadly follows the X-ray flux, and $N_{\rm diskbb}$ varies in the opposite direction (Fig.~\ref{fig:hht_fit}). 
The fitted $\Gamma$ and covering fraction vary less clearly with phase. 
A similar behavior was reported by \citet{fan2025nicer} for the heartbeat state observed in 2023, based on phase folding over a representative cycle. 
They found a positive correlation between flux and $T_{\rm in}$ and an anticorrelation between flux and the \texttt{diskbb} normalization, while the coronal parameters showed no significant correlation.

The phase-resolved behavior is compatible with a radiation-pressure-driven thermal--viscous cycle in the inner disk \citep{lightman1974black,belloni1997unified,done2007modelling,neilsen2011physics,neilsen2012radiation}. 
In such a cycle $T_{\rm in}$ and $N_{\rm diskbb}$ change during modulation.  
The correlation between flux and $T_{\rm in}$, together with the anticorrelation between flux and $N_{\rm diskbb}$, as shown in Table~\ref{tab:relation}, matches this disk-dominated picture.  
The 2025 event, however, has a much lower rms amplitude and a less regular light-curve morphology than the heartbeat state observed in 2023. 
This could occur if only part of the inner disk participates in the modulation, with the variable disk emission diluted by a more stable disk--corona component \citep{fan2025nicer,uttley2025}. 
It could also indicate that the disk approaches the large-amplitude instability regime without developing a full heartbeat cycle.

This interpretation remains consistent with the diverse phenomenology of millihertz-scale variability observed in BHXRBs, where heartbeat-like oscillations can differ substantially in amplitude and recurrence time across systems. 
IGR~J17091--3624 shows heartbeat-like flares with a broad range of recurrence times and fractional rms amplitudes \citep{altamirano2011faint}. 
HHT phase-resolved spectroscopy of its Class X heartbeat-like variability shows strong disk-temperature and disk-flux variations over the flare cycle, together with changes in the Comptonized component \citep{shui2024phase}. 
The QRM observed in 2025 in 4U~1630--47 is weaker and less regular than these strong heartbeat-like events, but its disk-dominated phase dependence is qualitatively similar.

The QRM observed in 2025 should also be distinguished from other millihertz-scale QPOs that may have different physical origins. 
In H1743--322, an unusual $\sim11$ mHz QPO was detected only in a small number of observations during the early outburst rise and was argued to be distinct from both type-C QPOs and classical heartbeat oscillations \citep{altamirano2012low}. 
A later phase-resolved study found evidence of iron-line flux modulation at twice the millihertz-scale QPO frequency, which was interpreted  in terms of Lense--Thirring precession of the inner flow \citep{cheng2019phase}. 
Although the disk parameters in H1743--322 show phase-dependent trends similar to those observed in 4U~1630--47, the most distinctive feature in H1743--322 is the modulation of the iron-line flux (see Figs.~3 and 6 of \citealt{cheng2019phase}), whereas the QRM observed in 2025 in 4U~1630--47 is primarily characterized by systematic changes in the disk parameters.

\section{Conclusions}
\label{conclusion}
We analyzed the NICER observations of 4U~1630--47 during its 2025 outburst, focusing on the LFQPOs in the rising phase and on the weak mHz QRM detected near the outburst peak. The main results are summarized below.

1. During the rising phase, the LFQPO centroid frequency increased from $\sim0.24$~Hz to $\sim3.43$~Hz as the source brightened and softened. Wavelet analysis shows that the LFQPO is intermittent within individual observations. Compared with the without-QPO intervals, the QPO intervals are generally hotter and have a lower apparent $N_{\rm diskbb}$; the changes in $\Gamma$ and covering fraction are smaller and less systematic. These differences indicate that the QPO intermittency  is  associated with changes in the inner accretion structure.

2. Near the peak of the outburst, we detect a weak mHz QRM at $\sim0.07$~Hz with a fractional rms of $\sim4.7\%$. This signal is weaker and less regular than the heartbeat state observed in 2023. The HHT phase-resolved spectra show a similar qualitative behavior in the QRM observed in 2025 and the heartbeat state observed in 2023: in both cases, $T_{\rm in}$ follows the X-ray flux, $N_{\rm diskbb}$ varies in the opposite direction, and the photon index shows no significant phase dependence. The clearest spectral changes therefore appear in the disk-related parameters.

\begin{acknowledgements}
We thank the anonymous referee for constructive comments that helped improve the clarity, presentation, and scientific interpretation of the manuscript.
MM acknowledges the research
programme Athena with project number 184.034.002, which is (partly)
financed by the Dutch Research Council (NWO). 
Haifan~Zhu acknowledges support from the China Scholarship Council (CSC; Grant No.~202506270166). 
Wei Wang thanks the NSFC (12133007) and the National Key Research and Development Program of China (Grants No. 2021YFA0718503 and 2023YFA1607901) for support.
\end{acknowledgements} 

\bibliographystyle{aa}
\bibliography{refs}
\begin{appendix}
\section{Comparison of the frequency--$\Gamma$ relation with the 2021 outburst}
\label{app:gamam}
\begin{figure}[htbp]
\centering
    \includegraphics[width=0.8\columnwidth]{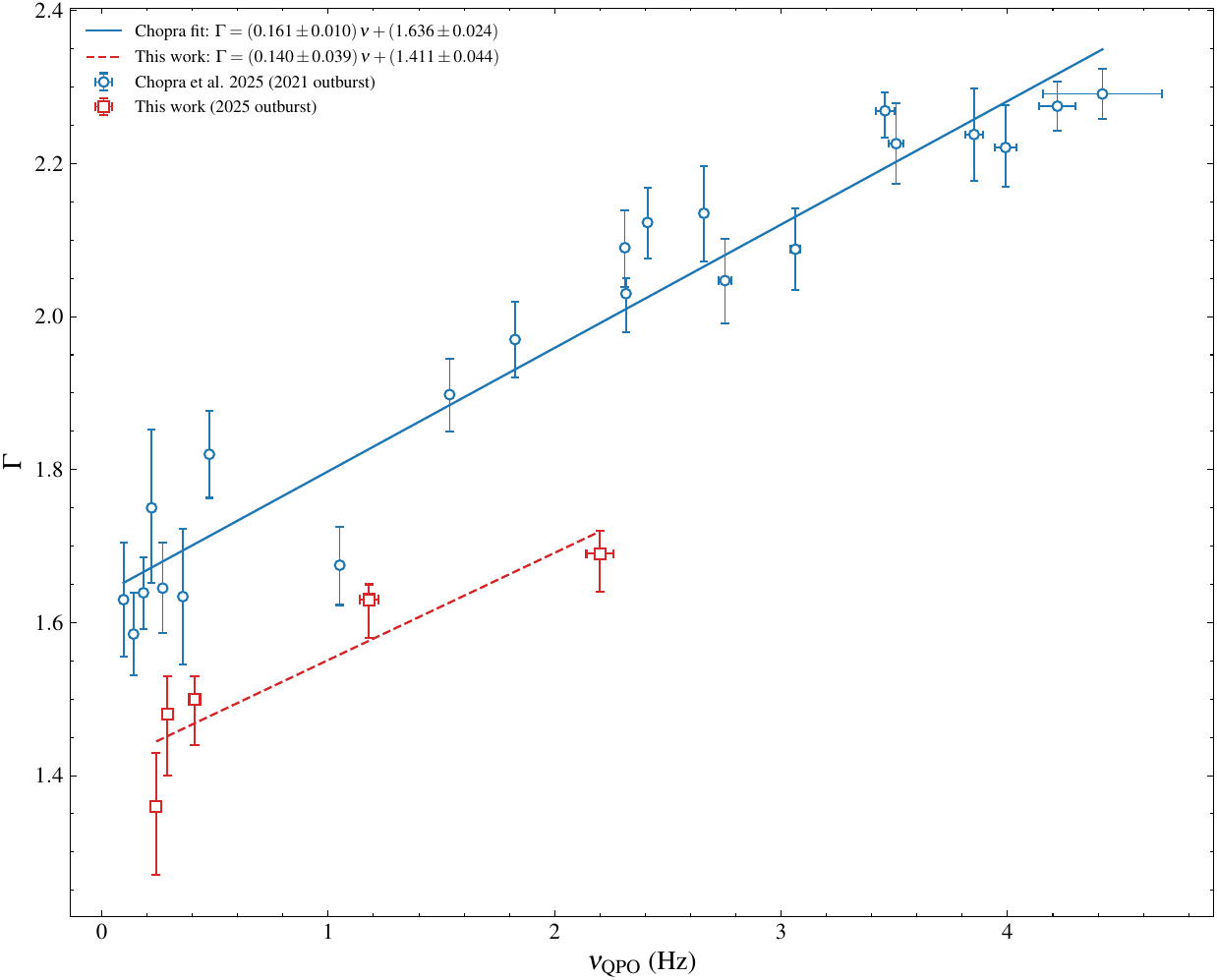}
\caption{
Comparison between the frequency--$\Gamma$ relation measured in this work for the 2025 outburst and that reported by \citet{chopra2025spectro} for the 2021 outburst. 
Only the QPO detections used for the long-term frequency--$\Gamma$ comparison are shown; the mHz QRM point is excluded. 
The dashed and solid lines show single linear fits to the 2025 and 2021 data, respectively. 
The two slopes are consistent within the uncertainties, but the 2025 points are systematically offset to lower $\Gamma$.
}
\label{fig:gamma_compare}
\end{figure}
To compare the long-term frequency--$\Gamma$ evolution between the two outbursts, we collected the type-C QPO measurements of \citet{chopra2025spectro} and plotted their centroid frequencies against the corresponding $\Gamma$. 
We included only the type-C QPO detections used in their frequency--$\Gamma$ analysis and excluded the later weak LFQPOs. 
For the 2025 outburst, we used the time-averaged spectral parameters from Table~\ref{tab:fitparams} and the QPO frequencies from Table~\ref{tab:qpo_params}, excluding the mHz QRM point. 
For simplicity, we did not adopt the two-slope fit used by \citet{chopra2025spectro} and instead fitted both samples with a single linear model.

Figure~\ref{fig:gamma_compare} shows that both outbursts exhibit a positive trend between $\nu_{\rm QPO}$ and $\Gamma$. 
A linear fit to the 2021 data gives $\Gamma=(0.161\pm0.010)\nu_{\rm QPO}+(1.636\pm0.024)$, while the 2025 data give $\Gamma=(0.140\pm0.039)\nu_{\rm QPO}+(1.411\pm0.044)$. 
The two slopes are consistent within the uncertainties, but the 2025 points are systematically offset toward lower $\Gamma$.

\citet{chopra2025spectro} used the \texttt{TBABS*(POWERLAW+DISKBB)} model, in which $\Gamma$ describes the local slope of an empirical hard tail. 
In our model, \texttt{tbfeo $\times$ (thcomp $\otimes$ diskbb)}, the disk spectrum is partly Comptonized, and $\Gamma$ is coupled to the covering fraction and the disk parameters. 
Therefore, the two $\Gamma$ are not strictly equivalent.

\end{appendix}

\end{document}